\documentclass[pre,aps,twocolumn]{revtex4}

\usepackage{graphicx}
\usepackage{amsmath,empheq}
\usepackage{amssymb}
\usepackage{ifthen}
\usepackage{booktabs}
\usepackage{graphicx}
\usepackage{dcolumn}
\usepackage{bm}
\usepackage{longtable}
\usepackage{gensymb}

\usepackage{graphicx}
\usepackage{dcolumn}
\usepackage{bm}
\usepackage{longtable}
\usepackage{gensymb}
\usepackage{color}

\newcommand{\bb}{\begin{equation}}
\newcommand{\ee}{\end{equation}}
\newcommand{\ba}{\begin{eqnarray*}}
\newcommand{\ea}{\end{eqnarray*}}
\newcommand{\rhor}{\rho({\bf r})}
\newcommand{\dd}{{\rm d}}
\newcommand{\rr}{{\mathbf r}}
\newcommand{\dr}{{\rm d}{\bf r}}

\newcommand{\RR}{{\mathcal{R}}}

\begin{document}

\title{Meniscus osculation and adsorption on geometrically structured walls}

\author{Martin \surname{Posp\'\i\v sil}}
\affiliation{
{Department of Physical Chemistry, University of Chemical Technology Prague, Praha 6, 166 28, Czech Republic;}\\
 {The Czech Academy of Sciences, Institute of Chemical Process Fundamentals,  Department of Molecular Modelling, 165 02 Prague, Czech Republic}}                
\author{Andrew O. \surname{Parry}}
\affiliation{Department of Mathematics, Imperial College London, London SW7 2BZ, UK}
\author{Alexandr \surname{Malijevsk\'y}}
\affiliation{ {Department of Physical Chemistry, University of Chemical Technology Prague, Praha 6, 166 28, Czech Republic;}
 {The Czech Academy of Sciences, Institute of Chemical Process Fundamentals,  Department of Molecular Modelling, 165 02 Prague, Czech Republic}}

\begin{abstract}
\noindent We study the adsorption of simple fluids at smoothly structured, completely wet, walls and show that a meniscus osculation transition
occurs when the Laplace and geometrical radii of curvature of locally parabolic regions coincide. Macroscopically, the osculation transition is of
fractional, $7/2$, order and separates regimes in which the adsorption is microscopic, containing only a thin wetting layer, and mesoscopic, in which
a meniscus exists. We develop a scaling theory for the rounding of the transition due to thin wetting layers and derive critical exponent relations
that determine how the interfacial height scales with the geometrical radius of curvature. Connection with the general geometric construction
proposed by Rasc\'on and Parry is made. Our predictions are supported by a microscopic model density functional theory (DFT) for drying at a
sinusoidally shaped hard-wall where we confirm the order of the transition and also an exact sum-rule for the generalized contact theorem due to
Upton. We show that as bulk coexistence is approached the adsorption isotherm separates into three regimes: a pre-osculation regime where it is
microscopic, containing only a thin wetting layer, a mesoscopic regime, in which a meniscus sits within the troughs, and finally another microscopic
regime where the liquid-gas interface unbinds from the crests of the substrate.
\end{abstract}

\maketitle

\section{Introduction}
Capillary condensation \cite{nakanishi81, nakanishi83, evans84, evans85, evans86, hend, gelb}, wetting \cite{sullivan, dietrich, row, schick,
forgacs, bonn} and wedge-filling \cite{hauge, rejmer, wood99, abraham02, delfino, binder03, bernardino, our_prl, our_wedge} are but a few examples of
surface phase transitions in which the surface tension plays a crucial role determining the location of the phase boundary and even the nature of the
transition itself. Recent studies of capillary condensation in open slits, possessing corners or edges, have also highlighted another phase
transition involving the depinning of the meniscus, located at the end(s) of the capillary as the pressure is increased \cite{md1, md2}. At a
macroscopic level, meniscus depinning is continuous with the order of the transition depending on the wetting properties of the confining walls. For
example, it is third-order if the walls are completely wet and second-order if they are partially wet. In each case, the transition is rounded or
smoothed by a mesoscopic length scale, specifically the parallel correlation length, associated with wetting layers at the capillary walls. The
cross-over scaling describing the rounding of meniscus depinning shows a remarkable consistency with the underlying theory of wetting transitions,
including conjectured critical exponent relations, giving physically intuitive results for mesoscopic corrections to macroscopic predictions for the
adsorption at the phase boundary.

In this paper we point out that there is another type of phase transition involving a meniscus in a different confining geometry, which can be
understood macroscopically, which is also rounded by mesoscopic wetting length scales. We refer to this as meniscus osculation and is associated with
the vanishing of the meniscus, that is the change from macroscopic to microscopic adsorption, when its curvature coincides with the geometrical
curvature of a confining sculpted wall that is completely wet. As we shall show, macroscopically, meniscus osculation is continuous but of fractional
order, different to depinning, and that the rounding, while also due to wetting behaviour, is described by a different cross-over scaling theory.
This cross-over scaling will allow us to make connection with very general expectations of how substrate geometry influences wetting behaviour.

Our paper is arranged as follows. In the next section we recall briefly the macroscopic and cross-over scaling theory for meniscus depinning,
emphasizing the underlying consistency with the theory of (complete) wetting. We then turn to meniscus osculation beginning first with the
macroscopic description before discussing the rounding due to complete wetting layers. This allows us to make a connection with a more general
proposal for how geometry influences adsorption on structured surfaces \cite{nature}. In the second part of our paper, we present the results of a
microscopic model Density Functional Theory (DFT) study of adsorption at a corrugated, sinusoidal, wall which shows not only meniscus osculation but
also a modified complete wetting transition. Indeed, as the chemical potential approaches bulk coexistence, we show that the adsorption changes from
being determined by microscopic length scales (pre-osculation) to macroscopic (geometry dominated) length scales and back to microscopic length
scales when it eventually unbinds from the crests of the sinusoid.

\section{Meniscus Depinning and Meniscus Osculation}

Before describing the osculation transition, we recall the macroscopic and mesoscopic cross-over scaling theory for meniscus depinning in an open
capillary, following closely the presentation in Refs.~\cite{md1, md2}. This will act as a guide for developing and interpreting the scaling theory
for the osculation transition occurring in parabolic and sinusoidally  corrugated geometries.

 \subsection{Meniscus Depinning}

\subsubsection{Macroscopics}

\begin{figure}
\includegraphics[width=8cm]{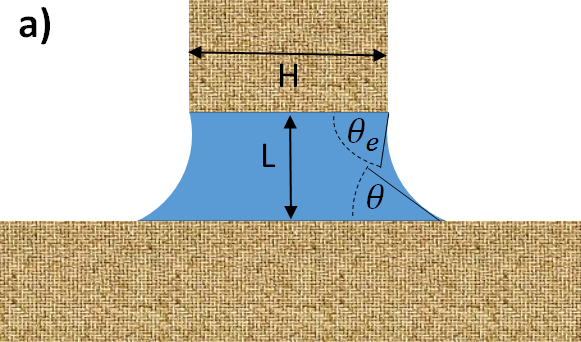}

\vspace*{1cm}

\includegraphics[width=8cm]{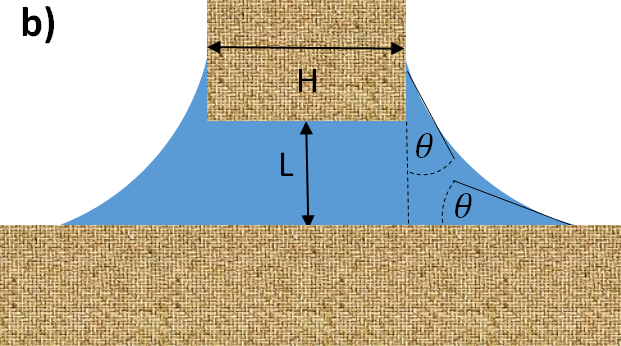}
\caption{Schematic illustration of two possible condensed capillary liquid phases in an open slit. In the top panel (a) the two circular menisci of
the Laplace radius of curvature are pinned at the upper edges which they meet at an edge contact angle $\theta_e$. The bottom of the menisci meets
the horizontal, lower, wall at the equilibrium contact angle $\theta$. In the lower panel (b) the two circular menisci are unpinned, spilling out
into the right-angle corners and meet both the vertical and lower walls at the contact angle $\theta$.}
\end{figure}

Consider an open capillary, made from two opposing planar walls separated by a distance $L$. The bottom wall is of infinite extent while the top wall
is of finite length $H$, which for our present purposes we suppose is much larger than $L$. Each end of the capillary is a right-angle edge and is in
contact with a bulk reservoir of gas at temperature $T$, far below the bulk critical temperature $T_c$, and pressure $p$ (or chemical potential
$\mu$) below bulk saturation $p_{\rm sat}$ (see Fig.~1). We denote $\delta p=p_{\rm sat}-p$. All the walls are considered completely wet by the
liquid phase corresponding to an equilibrium contact angle $\theta=0$. We suppose that the pressure is sufficiently close to bulk saturation that the
fluid within the capillary has already condensed to a liquid-like phase. For very long capillaries, this capillary condensation will occur near
$\delta p_{\rm cc} \approx 2\gamma/L$, as dictated by the macroscopic Kelvin equation, where $\gamma$ is the surface tension of the liquid-gas
interface. The full $H$ dependence of the pressure shift $\delta p_{\rm cc}$ can be determined but is not needed for the present purposes \cite{md1,
md2}. The capillary liquid phase is characterized by two menisci each of which are arcs of circles of Laplace radius $R=\gamma/\delta p$ located near
the ends of the capillary. Since the equilibrium contact angle $\theta=0$ these meet the bottom wall tangentially. However, there are two
possibilities for how they attach to the top wall. For $\delta p > \delta p_{\rm md}$, with $\delta p_{\rm md}=\gamma/L$, the menisci are pinned at
the corners and are characterized by an edge contact angle $\theta_e$. This is pressure dependent, given by
\begin{equation}
 \theta_e=\cos^{-1}\left(\frac{L-R}{R}\right)\,.
\end{equation}
 As the pressure is increased, the edge contact angle
increases until at $\delta p=\delta p_{\rm md}$, equivalent to $R=L$, it reaches its maximum value $\theta_e=\pi/2$, at which point it is tangential
to the vertical sides of the upper wall and hence the menisci are no longer pinned. Clearly, at the point of this depinning the menisci are each
quarter circles which just fit inside the open ends of the capillary. For $\delta p<\delta p_{\rm md}$ the menisci are unpinned and meet the side
walls tangentially at a distance $R-L$ above the corners. For complete wetting, meniscus depinning is a continuous third-order phase transition. This
is most easily seen by determining the excess adsorption $\Delta\Gamma$ beyond the contribution $\Delta\rho HL$ from the volume of liquid within the
capillary, with $\Delta\rho$ the difference in the bulk liquid and gas densities. For the pinned phase, with $R<L$ this is given by
\begin{equation}
 \Gamma= 2 \Delta\rho R^2 \left[\sin \theta_e +\frac{\sin 2\theta_e}{4}+\frac{1}{2}(\theta_e-\pi)\right]\,, \label{Gammapincomplete}
 \end{equation}
 while in the unpinned phase, for which $R>L$, it is given, trivially, by
\begin{equation}
 \Gamma = 2\Delta\rho \left(1-\frac{\pi}{4}\right)R^2\,. \label{gamma_dep_trans}
 \end{equation}
  It is then easy to check that
the second derivative of the adsorption is discontinuous at the depinning transition satisfying
 \bb
 \frac{\partial^2\Gamma}{\partial R^2}=\left\{\begin{array}{cc}
 \Delta\rho(4-\pi)\,;&\delta p=\delta p_{md}^-\,,\\
\Delta\rho(2-\pi)\,;&\delta p=\delta p_{md}^+\,. \label{delta_gam}
\end{array}\right.
 \ee
This is associated with a discontinuity in the third-derivative of the grand potential (per unit length of the capillary) showing that the transition
is indeed third-order, at the present macroscopic level.

\subsubsection{Mesoscopics}

The third-order meniscus depinning transition is rounded by length scales associated with complete wetting layers which coat the bottom and side
walls. The mechanism for this rounding is different to the standard finite-size scaling theory of the rounding of phase transitions when the
available volume is finite (or the dimensionality is below that of the lower critical dimension) which is due to fluctuations \cite{barber, privman}.
The rounding of the meniscus depinning transition is not due to fluctuation effects but rather due to mesoscopic length scales which smooth the
transition from pinned to unpinned states. Meniscus depinning is not associated with coexisting phases, symmetry breaking or a diverging
order-parameter and therefore there cannot be any true critical behaviour. The rounding of this transition is present as soon as we go beyond
macroscopics, for example, in mean-field DFT or even simpler interfacial models which accommodate the presence of wetting layers. The length scales
of relevance here are the wetting layer thickness $\ell_\pi \sim \delta p^{-\beta_s^{\rm co}}$ and parallel correlation length $\xi_\parallel \sim
\delta p^{-\nu_\parallel^{\rm co}}$, arising from interfacial fluctuations, which characterize wetting layers at planar walls, close to bulk
saturation. We note that the critical exponents, depend on the range of the intermolecular forces, but always satisfy the exact exponent relation
\cite{sullivan, dietrich, row, schick, forgacs, bonn}
\begin{equation}
 1+\beta_s^{\rm co}=2\nu_\parallel^{\rm co}\,, \label{exprel}
 \end{equation}
which is of relevance to the cross-over scaling theory for the rounding of the meniscus depinning transition. This exponent relation is the
consequence of an exact sum rule which determines that for wetting transitions $\partial\Gamma/\partial\mu\propto \xi_\parallel^2$
\cite{evans_tar84}. As described above, macroscopically, the depinning transition occurs when $R=L$ . Certainly, we can anticipate that the presence
of the wetting layer at the bottom wall decreases the effective width of the slit by $\ell_\pi$. However, of greater relevance is the rounding
arising from the parallel correlation $\xi_\parallel$, associated with the wetting layer at the vertical side walls, which smooths the point of
contact of the upper part of the meniscus with the corner (see Fig.~2). Since $\xi_\parallel$ is larger than $\ell_\pi$ we can therefore expect that
the depinning begins to occur when
 \begin{equation}
 R-L \approx \xi_\parallel\,. \label{RminL}
 \end{equation}
 This simple, finite-size scaling, observation is
the basis for the cross-over theory. The purely macroscopic result (\ref{delta_gam}) implies that the grand potential contains a singular
contribution $\Omega_{\rm sing}=\gamma (R-L)^3/6L^2$. To allow for the rounding due to complete wetting layers we modify this by a multiplicative
scaling function $W_{\rm md}(x)$ whose argument is the dimensionless scaling variable $x=(R-L)/\xi_\parallel$. Thus, the appropriate scaling ansatz
is
\begin{equation}
\Omega_{\rm sing}= \frac{(R-L)^3}{L^2}W_{\rm md} \left(\frac{R-L}{L^{\nu_\parallel^{\rm co}}}\right)\,,
\end{equation}
where we have ignored all constants and metric factors and substituted that, near the depinning transition, the correlation length  scales with the
slit width as $\xi_\parallel \approx L^{\nu_\parallel^{\rm co}}$.  We require that $W_{\rm md}(x)\to 0$ as $x\to\infty$ and $W(x)\to 1$ as
$x\to-\infty$ which represent the macroscopic unpinned and pinned states, respectively. The form of $W_{\rm md}(x)$ describes the smooth crossover
between these two states when the mesoscopic wetting length-scale $\xi_\parallel$ is allowed for. In order that $\Omega_{\rm sing}$ is non-zero at
the macroscopic meniscus depinning transition, $R=L$, we require that $W_{\rm md}(x)\propto 1/x^3$, which leads to a singular or mesoscopic
contribution. The derivative of $\Omega_{\rm sing}$ w.r.t. $\delta p$ determines the singular contribution to the adsorption, over and above the
macroscopic contribution $\Gamma \propto L^2$. Since $\partial\Omega/\partial \delta p \propto R^2\partial\Omega/\partial R$ it follows that the
adsorption contains a singular contribution
\begin{equation}
\Gamma_{\rm sing}= (R-L)^2\Lambda_{\rm md}\left(\frac{R-L}{L^{\nu_\parallel^{\rm co}}}\right)\,,
\end{equation}
where $\Lambda_{\rm md}(x)$ is a suitable new scaling function. We require that $\Lambda_{\rm md}(x)\to 0$ as $x\to\infty$, $\Lambda_{\rm md}(x)\to
1$ as $x\to-\infty$ and $\Lambda_{\rm md}(x)\propto 1/x^2$ as $x\to0$. It follows that exactly at the (macroscopic) depinning phase boundary, $R=L$,
the excess adsorption contains a mesoscopic contribution $ \Gamma_{\rm sing} \propto L^{2\nu_\parallel^{\rm co}}$ in addition to the leading-order
macroscopic term $\Gamma=2\Delta \rho \left(1-\frac{\pi}{4}\right) L^2$ determined earlier. Using the exact exponent relation (\ref{exprel}) it
follows that the mesoscopic term can be written
\begin{equation}
\Gamma_{\rm sing}\propto \ell_\pi L\,, \hspace{1cm} R=L\,,
\end{equation}
which is, of course, simply the additional contribution to the adsorption from the meniscus when we shift its position by the thickness of a wetting
layer, $\ell_\pi\sim L^{\beta_s^{\rm co}}$ coating the side walls (see Fig.~2). Explicitly, for systems with dispersion forces this yields
$\Gamma_{\rm sing} \propto L^{4/3}$.

\begin{figure}
\includegraphics[width=6cm]{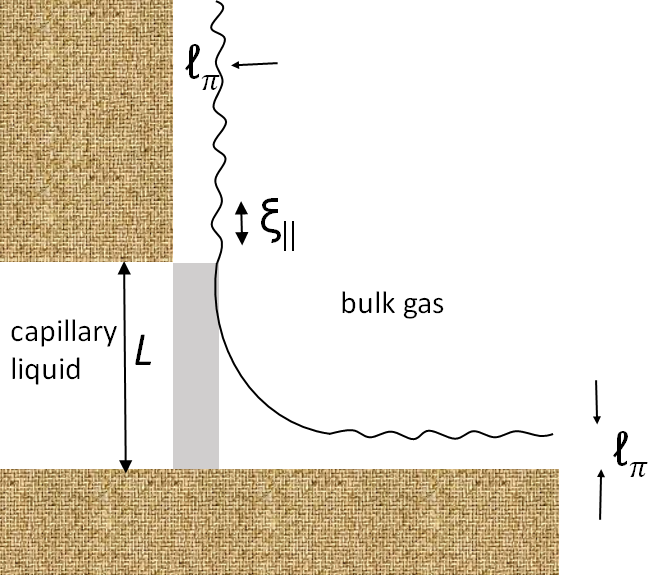}
\caption{Schematic illustration of the length scales determining the mesoscopic rounding of the continuous meniscus depinning transition, due to
wetting layers adsorbed along the bottom and side walls. The macroscopic phase boundary $R=L$ for meniscus depinning is rounded by the wetting layer
thickness $\ell_\pi$ and parallel correlation length $\xi_\parallel$ along the bottom and top walls, respectively. The gray area depicts an
additional contribution to the excess adsorption $\Gamma\propto \ell_\pi L$ due to the presence of the wetting layers. } \label{meso}
\end{figure}

Before moving on to consider the meniscus osculation transition we point out that the above cross-over scaling theory has a remarkable internal
consistency which appears to connect, very naturally, the macroscopics of the meniscus with the microscopic physics associated with wetting
behaviour. Suppose, for example we did not know what length scale was responsible for rounding the depinning transition. Instead, we simply marry the
macroscopic result that $\Gamma\approx (R-L)^2$ with the direct geometrical requirement that at $R=L$ there must be a higher order contribution $L
\ell_\pi$, due to the wetting layers at the side walls which just shift the position of the meniscus. To do this we, we write $\Gamma_{\rm sing}=
(R-L)^2\Lambda_{\rm md}({(R-L)}/\lambda_{\rm md})$ which contains an, as yet, undetermined length scale $\lambda_{\rm md}$ responsible for the
rounding. At $R=L$, it follows that the singular contribution must scale as $\lambda_{\rm md}^2$ which we identify with $L \ell_\pi$. Substituting
that $\ell_\pi \sim \delta p^{-\beta_s^{\rm co}}$ and that $R=\gamma/\delta p$ it follows that the length scale responsible for the rounding, must
behave as $\lambda_{\rm md}\sim \delta p^{-(1+\beta_s^{\rm co})/2}$. Using the exact exponent relation (\ref{exprel}) it follows that length scale
responsible for the rounding must be
 \bb
 \lambda_{\rm md}=\xi_\parallel\,,
 \ee
 consistent with Eq.~(\ref{RminL}).
 This is a remarkable self consistency -- adding a geometrical shift to the adsorption, due to the wetting layer thickness
$\ell_\pi$, we have generated a length scale, associated with wetting, which is the parallel correlation length. In other words, crossover scaling
demands that the exponent relation $1 + \beta_s^{\rm co}=2\nu_\parallel^{\rm co}$ is true.

 \subsection{Osculation transition} \label{osculation}

 \subsubsection{Macroscopics}

  \begin{figure}
\includegraphics[width=8cm]{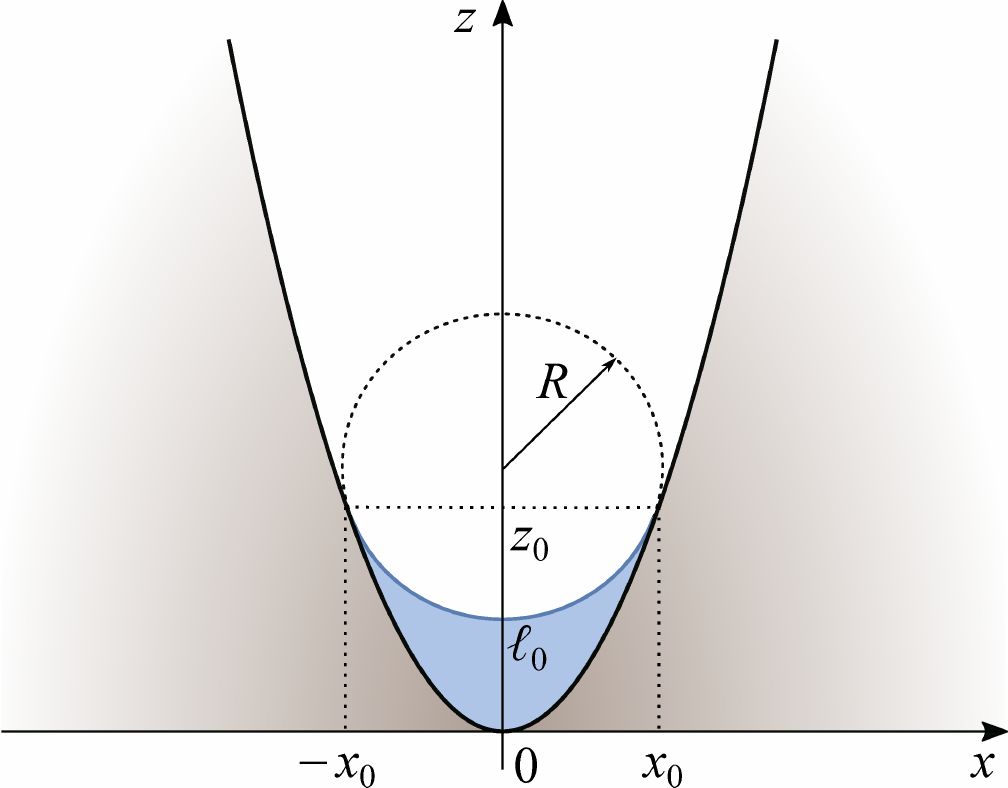}
\caption{Geometrical construction determining the macroscopic meniscus of in a parabolic, completely wet wall. The dashed circle is of
$R=\gamma/\delta p$ and connects with the wall tangentially at points $(-x_0,z_0)$ and $(x_0,z_0)$. The height of the meniscus above the wall is
$\ell_0$. The construction is only possible if $R$ is larger than the geometrical radius of curvature at the bottom., $R_w$. } \label{circle}
\end{figure}

We now turn attention to the phenomena of meniscus osculation and suppose that the confining wall has a cross-section of a parabola
 \begin{equation}
 \psi(x)=\frac{x^2}{2 R_w}\,,
 \end{equation}
with translational invariance assumed in the other direction. We restrict ourselves entirely to walls that are completely wet corresponding to
equilibrium contact angle $\theta=0$. The wall is again supposed to be in contact with a bulk gas at subcritical temperature $T$ and pressure
$p<p_{\rm sat}$. As the pressure is increased to bulk coexistence the adsorption diverges due to the growth of a meniscus near the bottom of the
parabola. At a macroscopic level this once again takes the shape of a circular arc of Laplace radius $R=\gamma/\delta p$ which meets the walls
tangentially at $\pm x_0$ at height $z_0$. We denote by $\ell_0$ the height of the meniscus above the bottom (see Fig.~3). These length scales are
trivially determined yielding
 \begin{eqnarray}
x_0&=&\sqrt{R^2-R_w^2}\,, \label{x0}\\
z_0&=&\frac{R^2-R_w^2}{2R_w}\,, \label{z0}\\
\ell_0&=&\frac{(R-R_w)^2}{2R_w}\,.  \label{l0}
 \end{eqnarray}

As pointed out in Ref.~\cite{nature}, as $\delta p \to 0$ these length scales diverge with universal exponents which are geometry-dominated,
independent of the range of the intermolecular forces. More generally, for walls which have a power-law cross-section $\psi(x)\propto |x|^\phi$, the
divergence of the adsorption as $\delta p\to0$ is geometry dominated provided that $\phi>\beta_s^{\rm co}/\nu_\parallel^{\rm co}$. For
$\phi<\beta_s^{\rm co}/\nu_\parallel^{\rm co}$ the adsorption diverges similar to complete wetting at a planar wall which depends strongly on the
range of intermolecular forces. Returning to the parabola, we focus here on the \emph{disappearance} of the meniscus as the pressure is reduced to
point at which
 \begin{equation}
  R=R_w\,.
 \end{equation}
We refer to this as meniscus osculation. Macroscopically, for $R<R_w$ there is no adsorption of liquid at the walls since no meniscus can be fitted
into the geometry. More generally, beyond macroscopics, meniscus osculation separates regimes in which the adsorption is geometry-dominated and
microscopic, respectively. Associated with the osculation is a singular contribution to the grand potential $\Omega$ per unit length of the wall. At
a purely macroscopic level this is determined by
 \bb
  \Omega=\delta pA+\gamma(\ell_m-\ell_w)\,, \label{omega}
 \ee
where $A$ is the area of liquid, $\ell_m$ is the meniscus length and $\ell_w$ is contact length of the liquid with the wall. Again, these are very
simply determined as
\begin{eqnarray}
A&=&\frac{x_0}{3R_w}(2R^2+R_w^2)-R^2\sin^{-1}\left(\frac{x_0}{R}\right)\,, \label{A}
 \end{eqnarray}
\bb
 \ell_m=2\sin^{-1}\left(\frac{\sqrt{R^2-R_w^2}}{R}\right) \label{lm}
 \ee
 and
\bb
 \ell_w=\frac{R}{R_w}\sqrt{R^2-R_w^2}+R_w\sinh^{-1}\left(\frac{\sqrt{R^2-R_w^2}}{R_w}\right)\,. \label{lw}
 \ee
This implies that the grand potential is given explicitly by
  \begin{eqnarray}
  \frac{\Omega}{\gamma}&=&\frac{\sqrt{R^2-R_w^2}}{3R_w}\left(\frac{R_w^2}{R}-R\right)+R\sin^{-1}\left(\frac{\sqrt{R^2-R_w^2}}{R}\right)\nonumber\\
       &&-R_w\sinh^{-1}\left(\frac{\sqrt{R^2-R_w^2}}{R_w}\right)\,.
 \end{eqnarray}
As the pressure is decreased, and $R$ reduces to $R_w$, this exhibits the (macroscopic) critical singularity
 \bb
\frac{\Omega}{\gamma}\approx-\frac{16\sqrt{2}}{105}\frac{(R-R_w)^{\frac{7}{2}}}{R_w^\frac{5}{2}}\,, \label{sevenhalf}
 \ee
which is of fractional, higher-order, than the third-order meniscus depinning. The derivative of the grand potential determines the associated
singularity in the adsorption which disappears as
 \bb
\Gamma_{\rm sing} \approx\frac{8\sqrt{2}}{15}\frac{\Delta\rho (R-R_w)^\frac{5}{2}}{\sqrt{R_w}}\,, \label{gamma_osc_macro}
 \ee
which is simply $\Delta\rho$ times the meniscus area $A$. For $R<R_w$ there is no macroscopic adsorption.

Meniscus osculation is closely related to the phenomena of capillary emptying in horizontal capillaries, under the influence of gravity, of
elliptical cross-section \cite{empty}. Specifically, the rich phase behaviour associated with emptying is related to whether menisci, which now
represent the cross-sectional shape of horizontal liquid tongues, can be locally inscribed within the ellipse -- see e.g. the cross-sections shown in
Figs.~1 and 2 in Ref.~\cite{empty}.

\subsubsection{Mesoscopics}

Similar to depinning, the continuous osculation transition must be rounded by length scales due to mesoscopic complete wetting layers. There will
always be some residual microscopic adsorption even when $R<R_w$ implying that the cross-over from the geometry-dominated regime, $R>R_w$, must be
smooth since there is neither symmetry-breaking nor any diverging order parameter or length scale. For the meniscus depinning the rounding of the
macroscopic phase boundary $R=L$ can be understood directly, by considering how the macroscopic meniscus attaches to the edge of the opening when
wetting layers are present, leading to $R-L\approx \xi_\parallel$. This line of reasoning is not so obvious for the rounding at osculation since the
meniscus itself is disappearing. The simplest length scale and observable to focus on is the value of the interfacial height $\ell_0$ when $R=R_w$
for which there is no macroscopic contribution. We wish to determine how, exactly at osculation, $R=R_w$, the interfacial height scales with radius
of curvature and introduce a new osculation critical exponent, $\beta_{\rm osc}$, to characterise this;
 \bb
  \ell_0\approx R_w^{\beta_{\rm osc}}\,,\hspace{1cm} R=R_w\,,
 \ee
where we have omitted any unimportant dimensional pre-factors. To determine the osculation exponent, we suppose that in the vicinity of the
osculation transition the interfacial height shows crossover scaling
 \bb
 \ell_0=\frac{(R-R_w)^2}{2R_w}\mathcal{L}_{\rm osc}\left(\frac{R-R_w}{\lambda_{\rm osc}}\right)\,, \label{ell_0_meso}
 \ee
with $\lambda_{\rm osc}$ the undetermined rounding length scale and $\mathcal {L}_{\rm osc}(x)$ the appropriate scaling function. The macroscopic
limit, corresponding to Eq.~(\ref{l0}), is recovered by imposing that  $\mathcal {L}_{\rm osc}(x)\to1$, as $x\to\infty$. We also require that
$\mathcal{L}_{\rm osc}(x) \sim x^{-2}$, as $x\to0$,  in order that at osculation the value of $\ell_0$ remains finite implying $\ell_0 \propto
\lambda_{\rm osc}^2/R_w$.  Since the wall is completely wet we also require that this diverges as $R_w \to\infty$, i.e. as we follow the line of
osculation transitions to bulk coexistence. This means that we can eliminate the possibility that the rounding length scale is the planar wetting
layer thickness $\ell_\pi$, since in that case $\ell_0$ does not diverge. In addition, the identification $\lambda_{\rm osc}\approx\xi_\parallel$,
while appropriate for meniscus depinning, is unsatisfactory since, on using the exact exponent relation (\ref{exprel}), this only yields
$\ell_0\approx \ell_\pi$. This is most likely only a lower bound since it is physically reasonable that the residual interfacial height at osculation
is greater than the planar wetting layer thickness at the same pressure.

To identify the rounding length scale we instead focus on the value of $\ell_0$ deep in the pre-osculation regime, $R_w\gg R$, and ensure that our
scaling ansatz is compatible with the necessary curvature induced enhancement of the adsorption. For wetting on spheres and cylinders of radius $R_c$
(say), which have a negative curvature, it is well known that the increase in the surface tension contribution thins the wetting layer equivalent to
a shift in the effective value of the partial pressure from $\delta p$ to $\delta p + \gamma/R_c$ \cite{holyst, bieker, stewart, morgan, nold}. For
the parabola it is therefore reasonable to expect that when $R_w\gg R$ the local interface height is enhanced by the positive curvature so that
$\ell_0$ is the same as the planar wetting layer thickness but evaluated at an effective reduced partial pressure $\delta p-\gamma/R_w$, i.e.
$\ell_0\to (1/R-1/R_w)^{-\beta_s^{co}}$. This requirement is highly restrictive and is only compatible with the crossover scaling ansatz
(\ref{ell_0_meso}) provided that the scaling function $\mathcal{L}_{\rm osc}(x)\propto |x|^{-2-\beta_s^{co}}$, as $x\to-\infty$, and identifies that
the rounding length-scale scales with $R_w$ and $R$ according to
\begin{equation}
\lambda_{\rm osc}^{2+\beta_s^{\rm co}}= R_w^{1+ \beta_s^{\rm co}}R^{\beta_s^{\rm co}}
\end{equation}
where again we have omitted unimportant microscopic length scales. Using this value for $\lambda_{\rm osc}$ it follows that the osculation and
wetting exponents are related via
\begin{equation}
\beta_{\rm osc}= \frac{3 \beta_s^{\rm co}}{2+\beta_s^{\rm co}}
\end{equation}
which is the central prediction of our mesoscopic scaling theory. In 3D and with dispersion forces, and also in 2D with short-ranged forces, for
which $\beta_s^{\rm co}=1/3$ this predicts that $\beta_{\rm osc}=3/7$ pointing to a rather non trivial interplay between geometry and wetting at
osculation. In 3D and with short-ranged forces it is likely that the prediction $\beta_{\rm osc}=0$ corresponds to a logarithmic dependence of
$\ell_0$ on $R_w$.

\begin{figure*}
\includegraphics[width=17cm]{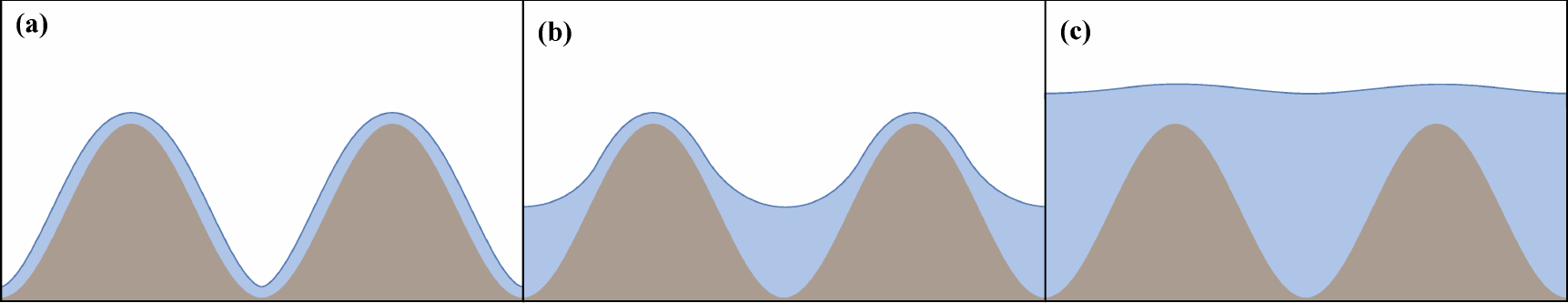}
\caption{A schematic illustration of three possible adsorption regimes of a completely wet ($\theta=0$) sinusoidal wall: At low-adsorption,
pre-osculation regime (a), far away from bulk coexistence, a microscopic amount of liquid coats the wall and the liquid-gas interface follows the
wall shape. Within the second regime (b), the voids of the wall become gradually filled with liquid. In the third regime (c), the liquid-gas
interface unbinds from the wall and its height diverges as the coexistence is approached, similarly, but not identically, to planar walls. While the
first and third regimes are governed by microscopic forces, the second regime is controlled by the wall geometry and can be described
macroscopically.} \label{sketch}
\end{figure*}

 \begin{figure}
\includegraphics[width=8cm]{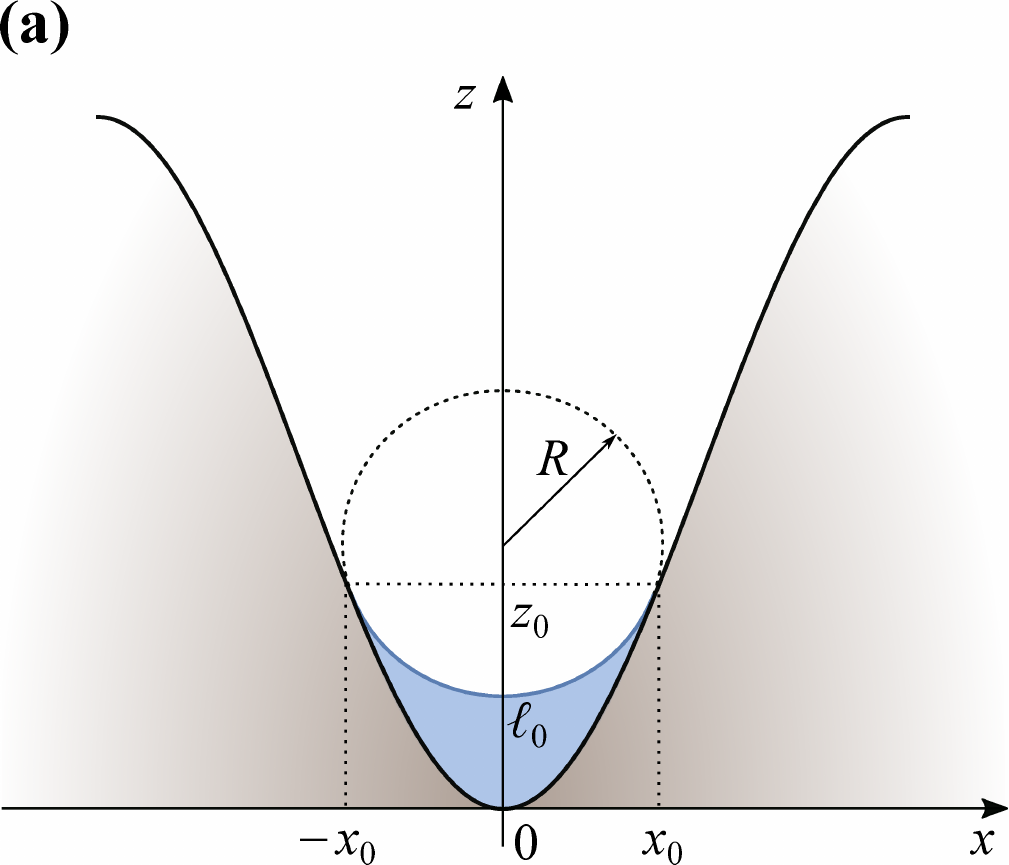} \\
\vspace{0.5cm}
\includegraphics[width=8cm]{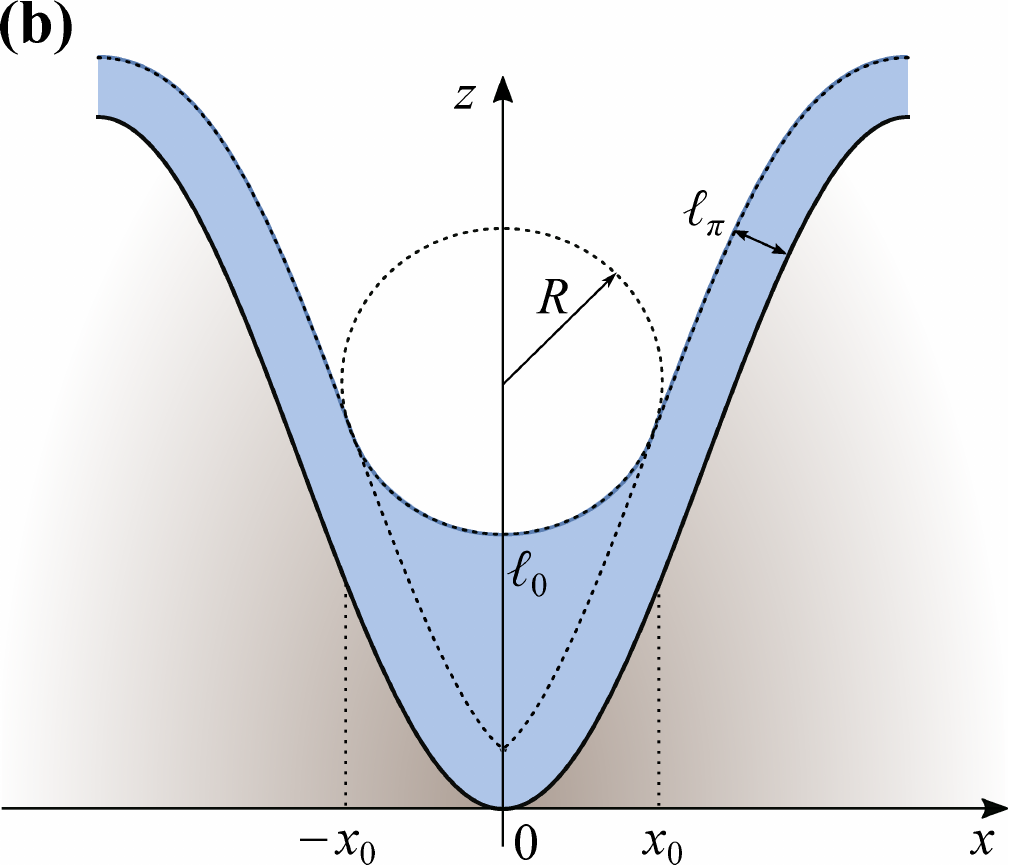}
\caption{Schematic illustration of (a) geometrical construction for macroscopic meniscus in a sinusoidal wall; (b) modified, RP construction where
the wall is first coated by a uniform layer of thickness $\ell_\pi$.} \label{circle}
\end{figure}

Having identified the rounding length scale $\lambda_{\rm osc}$ the crossover scaling theory can be applied to other quantities. For example, we can
expect that the singular contribution to the grand potential  per length scales as
\begin{equation}
\Omega =\frac{(R-R_w)^{7/2}}{R_w^{5/2}}W_{\rm osc} \left(\frac{R-R_w}{\lambda_{\rm osc}}\right)\,,
\end{equation}
where we have ignored the pre-factors.  The scaling function $W_{\rm osc}(x)$ must satisfy $W_{\rm osc}(\infty)=1$ in order to recover the
macroscopic singularity in the geometry-dominated regime. On approaching the osculation transition at $R=R_w$, the scaling function must also behave
as $W_{\rm osc}\propto x^{-7/2}$ in order to leave a finite, singular contribution, which we define as
\begin{equation}
\Omega\approx R_w^{2-\alpha_{\rm osc}}\,,
\end{equation}
This identifies the exponent as
\begin{equation}
2-\alpha_{\rm osc}=\frac{3}{2}\left(\frac{3\beta_s^{co}-1}{2+\beta_s^{co}}\right)\,,
\end{equation}
again, relating it to the singularities at complete wetting. This exponent relation is noteworthy because it predicts that $2-\alpha_{\rm osc}=0$
when $\beta_s^{co}=1/3$, as is pertinent to 3D systems with dispersion forces and 2D systems  with short-ranged forces, and is presumably indicative
of a marginal logarithmic singularity. We mention this because precisely the same marginal singularity is predicted for meniscus depinning for
systems with dispersion forces \cite{md1, md2}. Exactly at the depinning transition $R=L$ the singular contribution to the grand potential is
predicted to be $\Omega\approx L^{2-\alpha_{md}}$ where $2-\alpha_{md}=(3\beta_s^{co}-1)/2$ which obviously also vanishes when $\beta_s^{co}=1/3$.
The vanishing of both these exponents is consistent with the known marginal logarithmic contribution to the free-energy of a finite-size droplet in
these systems \cite{stripe17}. This consistency between osculation, depinning and complete wetting for systems with dispersion forces is not achieved
if we had chosen $\lambda_{\rm osc}\approx \xi_\parallel$ which, as mentioned above, we believe is a lower-bound for the rounding length scale at
osculation.

The prediction of the mesoscopic crossover scaling theory, that at osculation $\ell_\pi$ is a lower bound for the value of $\ell_0$, is consistent
with a simple idea proposed by Rasc\'on and Parry (RP) who suggested that the total adsorption on a completely wet sculpted surface can be determined
by first coating the wall with a wetting layer of thickness, $\ell_\pi$, defined normal to each surface point \cite{nature}. This coating modifies
the original wall shape, $\psi(x) \to \tilde \psi(x)$, on to which we then fit a meniscus of Laplace radius $R$. This simple modified macroscopic
construction reproduces with remarkable accuracy different scaling regimes, arising due to the competition between microscopic forces and geometry,
obtained from numerical studies of interfacial Hamiltonian models and experiments \cite{exp1, exp2, exp3}. We shall return to this in the next
section.


\section{Adsorption on a corrugated wall}

In preparation for studying meniscus osculation in a microscopic DFT we turn our attention to the adsorption occurring near a sinusoidally corrugated
wall described by the height function
 \bb
 \psi(x)=A\left[1-\cos\left(\frac{2\pi x}{L}\right)\right]\,, \label{psi}
 \ee
where $A$ is amplitude and $L$ is the period. The axes here are chosen such that $\psi=0$ at the bottom at wall. The adsorption and wetting
properties on such corrugated surfaces have been studied previously, but usually concentrating on the partial wetting regime where, for simple
fluids, pre-wetting and unbending transitions (equivalent to a local condensation of liquid within the troughs) compete \cite{rascon, rejmer2000,
kubalski2001, rejmer2002, rejmer2007, rodriguez}. The surface behaviour is very rich for complex fluids (liquid crystals) where dislocations near the
crests must be taken into account \cite{patricio1, patricio2}. Here we restrict ourselves to simple fluids and to complete wetting ($\theta=0$) where
none of these features are present. In spite of this, the adsorption isotherm is still quite rich and, we anticipate, falls into three regimes (see
Fig.~\ref{sketch}):

\subsection{Microscopic, pre-osculation regime} \label{regime1}

When the pressure is low, such that the Laplace radius of curvature $R\ll R_w$ where $R_w=L^2/4\pi^2 A$, no meniscus is present. In this case the
adsorption is entirely microscopic comprising a thin wetting layer of approximate thickness $\ell_\pi$ which coats the wall. As the pressure is
increased, a rounded meniscus osculation transition occurs near $R=R_w-\lambda_{\rm osc}$ where the adsorption changes from microscopic to
macroscopic due to the growth of the meniscus. In three dimensional systems and for short-range forces $\lambda_{\rm osc}\approx\sqrt{R_w}$ and
therefore is of a similar size to $\xi_\parallel$. A signature of this rounded transition would be a dramatic change in the behaviour of the second
and third derivatives, $\partial^2\Gamma/\partial\mu^2$ and $\partial^3\Gamma/\partial\mu^3$, which reflect the fractional order of the macroscopic
transition.

\subsection{Macroscopic, geometry-dominated regime} \label{regime2}

When $R<R_w$ a meniscus sits near the troughs of the corrugated wall connecting tangentially with its sides. As the pressure is increased the
meniscus grows until it is near the crests of the sinusoid. Macroscopically this, geometry-dominated, regime extends from osculation ($R=R_w$),
equivalent to
 \bb
\delta p_{\rm osc}=\frac{4 \pi^2 A\gamma}{L^2}\,,  \label{p_osc}
 \ee
to saturation, $\delta p=0$. It can again be shown easily that for the sinusoidal wall (\ref{psi}) the height of the meniscus above the bottom is
  \bb
 \ell_0=z_0+\sqrt{R^2-x_0^2}-R\,.   \label{ell0}
 \ee
 where $\pm x_0$ and $z_0$ are the coordinates denoting the contact of the meniscus with the wall (see Fig.~3) which are given implicitly
 by solving simultaneously:
  \bb
A\sin\left(\frac{2\pi x_0}{L}\right)=\frac{Lx_0}{2\pi\sqrt{R^2-x_0^2}} \label{sin1}
 \ee
and
 \bb
 z_0=A\left[1-\cos\left(\frac{2\pi x_0}{L}\right)\right]\,.  \label{sin2}
 \ee

The presence of wetting layers qualitatively and quantitatively affects this and is described approximately by the RP construction, in which we coat
the wall with the wetting layer prior to fitting inside the circular meniscus.

Similarly, the approach to saturation is more properly a cross-over from a geometry-dominated regime, where the meniscus lies within the wells, to a
second microscopic regime where the liquid-gas interface unbinds from the crests. This second cross-over occurs when the period
$L\approx\xi_\parallel$, i.e. when the characteristic scale of the interfacial fluctuations is similar to the geometric scale.

\subsection{Microscopic, interfacial unbinding regime} \label{regime3}

On approaching saturation, the liquid-gas interface must unbind from the wall and will display properties which are very similar to complete wetting
at a planar surface. The divergence of the film thickness in this limit cannot be described macroscopically and we must allow for the intermolecular
forces and the effective binding that results between the interface and the wall. The shape of the wall does have some influence on the equilibrium
shape of the liquid-gas interface, but this is negligible, at {\it{leading-order}}, when the parallel correlation length is larger than the period
$L$, corresponding to the regime $\delta p\ll L^{-1/\nu_\parallel^{\rm co}}$ in which case the interface is effectively planar. Critical exponents
which characterise the divergence of the film thickness and parallel correlation length are unchanged. The question whether there is any signature of
the wall shape on the complete wetting layer, as it unbinds from the crests, is rather subtle and is related to the nature of the binding potential.
This is particularly subtle for systems with short-ranged forces and we intend to focus on this problem in a separate work \cite{future}.

\section{DFT analysis of adsorption on sinusoidal hard wall}

Here we present the results for adsorption on sinusoidal hard wall obtained using a microscopic DFT. We present first the model whose accuracy we
test using an exact sum rule due to Upton \cite{upton}. Numerical results illustrating the meniscus osculation transition and different adsorption
regimes are then presented.

\subsection{Density functional theory}

Consider a simple fluid which is a subject of an external field $V(\rr)$ due to a presence of a confining wall and which is at contact with a bulk
reservoir at temperature $T$ and chemical potential $\mu$. Within classical DFT \cite{evans79} the equilibrium density profile $\rho(\rr)$ of the
fluid is obtained by minimisation of the grand potential functional
   \bb
  \Omega[\rho] =F[\rho]+\int\dr\rho(\rr)\left[V(\rr)-\mu\right]\,,\label{grandpot}
 \ee
where $F[\rho]$ is the intrinsic free energy functional containing all the information about the fluid interactions. Its approximation, which is a
crucial part of any DFT model, depends largely on the choice of the fluid model and  particularly for simple fluids the intrinsic free energy can be
treated in the following perturbative manner:
 \bb
 F[\rho]=F_{\rm id}[\rho]+F_{\rm hs}[\rho]+F_{\rm att}[\rho]\,, \label{f_dft}
 \ee
 which splits the functional into the ideal gas, $F_{\rm id}$, repulsive hard-sphere, $F_{\rm hs}$, and attractive, $F_{\rm att}$, contributions.

 The ideal gas part is known exactly and is given by
  \bb
  \beta F_{\rm id}[\rho]=\int\dr\rho(\rr)\left[\ln(\rhor\Lambda^3)-1\right]
  \ee
  where $\Lambda$ is the thermal de Broglie wavelength which can be set to unity and $\beta=1/k_BT$ is the inverse temperature.

The repulsive interaction between fluid molecules is approximated by the hard-sphere potential and its contribution to the free energy is described
accurately using Rosenfeld's fundamental measure theory \cite{ros}
 \bb
F_{\rm hs}[\rho]=\frac{1}{\beta}\int\dd\rr\,\Phi(\{n_\alpha\})\,,\label{fmt}
 \ee
 where $\{n_\alpha\}$ denotes a set of six weighted densities
 \bb
 n_\alpha(\rr)=\int\dr'\rho(\rr')\omega_\alpha(\rr-\rr')\,,\;\;\alpha=\{0,1,2,3,v1,v2\}\,, \label{n_alpha}
 \ee
given by convolutions between one-body fluid density $\rhor$ and the weight functions $\{\omega_\alpha\}$ which characterize so called fundamental
measures of the hard-sphere particles of diameter $\sigma$:
 \begin{eqnarray}
 \omega_3(\rr)&=&\Theta(\RR-|\rr|)\,,\;\;\;\;\omega_2(\rr)=\delta(\RR-|\rr|)\,,\\
 \omega_1(\rr)&=&\omega_2(\rr)/4\pi \RR\,,\;\;\;\,\omega_0(\rr)=\omega_2(\rr)/4\pi \RR^2\,,\\
 \omega_{v2}(\rr)&=&\frac{\rr}{\RR}\delta(\RR-|\rr|)\,,\omega_{v1}(\rr)=\omega_{v2}(\rr)/4\pi \RR\,.
 \end{eqnarray}
Here, $\Theta$ is the Heaviside function, $\delta$ is Dirac's delta function and $\RR=\sigma/2$. Among various possible prescriptions for describing
the free energy density $\Phi$ for inhomogeneous hard-sphere fluid, we adopt the original Rosenfeld approximation \cite{ros} which accurately
describes short-range correlations between fluid particles and satisfies exact statistical mechanical sum rules \cite{hend}.

For separations $r>\sigma$, a pair of fluid particles is assumed to interact via the attractive part of the Lennard-Jones potential, $u_{\rm
att}(r)$, which is truncated at a cut-off  which we set to be $r_c=2.5\,\sigma$, i.e.:
 \bb
 u_{\rm att}(r)=\left\{\begin{array}{cc}
 0\,;&r<\sigma\,,\\
-4\varepsilon\left(\frac{\sigma}{r}\right)^6\,;& \sigma<r<r_c\,,\\
0\,;&r>r_c\,.
\end{array}\right.\label{uatt}
 \ee
This attractive contribution is modelled in simple mean-field fashion:
 \bb
F_{\rm att}[\rho]=\frac{1}{2}\int\int\dd\rr\dd\rr'\rhor\rho(\rr')u_{\rm att}(|\rr-\rr'|)\,.
 \ee

We assume that the external potential is a hard wall of sinusoidal shape:
 \bb
 V(x,z) = \left\{
  \begin{array}{ll}
  \infty, & z< \psi(x)\,, \\
   0, & z > \psi(x)\,,
  \end{array} \right.
 \ee
where $\psi(x)$ is given by Eq.~(\ref{psi}). A purely hard is known to be completely dry, i.e. completely wet by vapour, corresponding to contact
angle $\theta=\pi$. By studying drying we are also able to avoid the complications arising from layering since liquid layers are absent near the
surface of the wall. We assume that the wall is of a macroscopic extent in the remaining Cartesian direction and the system is thus translationally
invariant along the $y$ axis.

The minimization of (\ref{grandpot}) leads to the self-consistent equation for the equilibrium density profile
  \bb
  \rho(\rr) = \Lambda^{-3} \exp\left[\beta\mu-\beta V(\rr) + c^{(1)}(\rr)\right]\,,  \label{selfconsistent}
 \ee
where $c^{(1)}(\rr)=c^{(1)}_\mathrm{hs}(\rr)+c^{(1)}_\mathrm{att}(\rr)$ is the one-body direct correlation function, which can be split into the
hard-sphere contribution
 \bb
  c^{(1)}_\mathrm{hs}(\rr)  
   = -\sum_\alpha \int\dd\rr'\; \frac{\partial\Phi(\{n_\alpha\})}{\partial n_\alpha} \, \omega_\alpha(\rr'-\rr)   \label{chs}
 \ee
and the attractive part
  \bb
  c^{(1)}_\mathrm{att}(\rr) 
  =-\beta\int \dd\rr'\; u_{\rm att}(|\rr-\rr'|)\,\rho(\rr')\,.\label{catt}
  \ee

We solve Eq.~(\ref{selfconsistent}) numerically on a rectangular discrete grid of spacing of $0.1\,\sigma$ using Picard iteration where the
convolutions in (\ref{n_alpha}), (\ref{chs}), and (\ref{catt}) are determined using a fast Fourier transform \cite{frink}.

From the equilibrium density profile $\rho(x,z)$ we can extract the local height of the liquid-gas interface identified as
 \bb
 \rho(x,\ell(x))=\frac{\rho_g+\rho_l}{2}
 \ee
 and also the excess adsorption
  \bb
  \Gamma=\int_{-L/2}^{L/2}\dd x\int_{\psi(x)}^\infty \dd z (\rho_b-\rho(x,z))\,, \label{ads}
  \ee
  where $\rho_b$ is the bulk density. In our numerical study, we are, of course, not able to model a semi-infinite system, instead, we set the
  density equal to the bulk density at a fixed distance of $50\,\sigma$ above the crests of the sinusoid.

  \begin{figure}
\includegraphics[width=8cm]{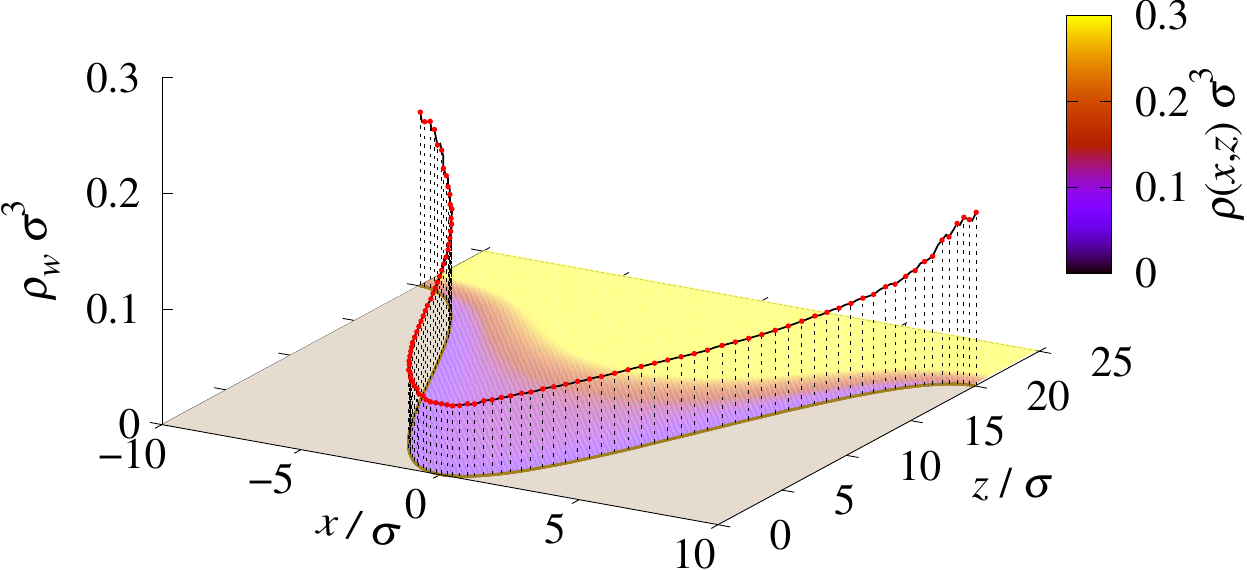}
\caption{Three-dimensional plot of the contact density profile for a  corrugated hard wall, (\ref{psi}), over a single period with $L=20\,\sigma$ and
$A=10\,\sigma$, which is in contact with an supersaturated bulk liquid at $\delta\mu=0.0023\,\varepsilon$. The black line with red dots represents
the value of the local contact density $\rho({\bf s})$ where ${\bf s}=(x,\psi(x))$. The corresponding full DFT density profile, in the form of a
colormap in the $xz$-plane, is also shown.} \label{integration}
\end{figure}

\begin{figure}[h]
\includegraphics[width=8cm]{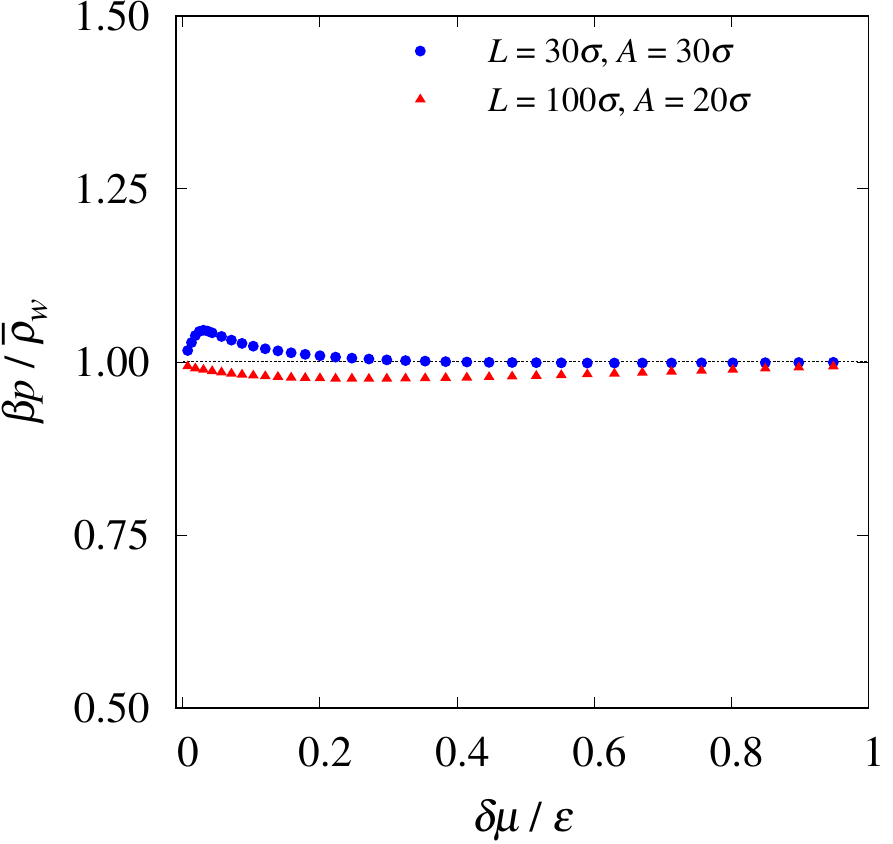}
\caption{Numerical DFT results for the dimensionless ratio $\beta p/\bar{\rho_w}$ as a function of the bulk fluid supersaturation $\delta\mu$  for
two different wall amplitudes and periods showing good agreement with Upton's generalized contact theorem (\ref{upton_t}).} \label{upton-fig}
\end{figure}

\begin{figure*}
\includegraphics[width=4.5cm]{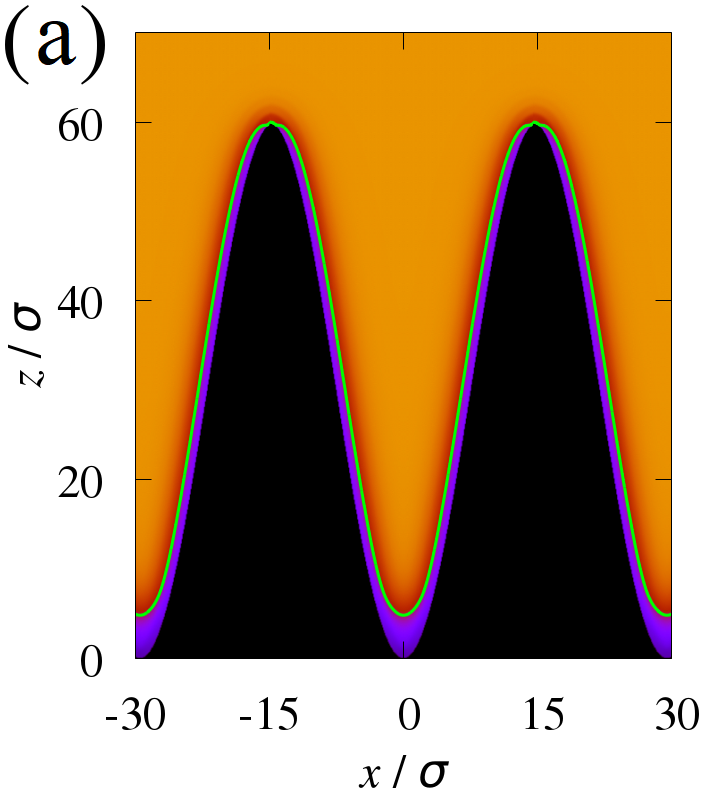} \hspace{0.5cm} \includegraphics[width=4.5cm]{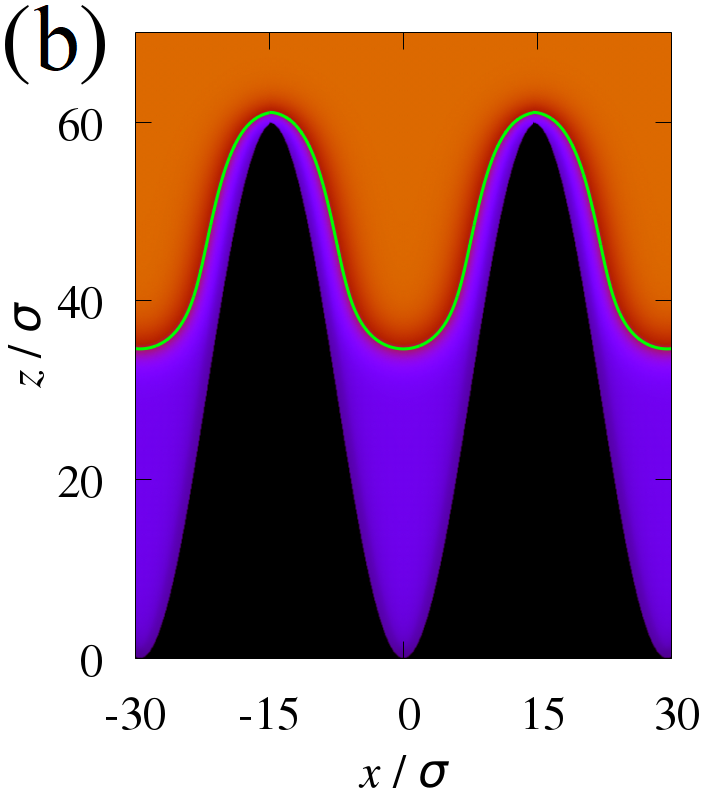} \hspace{0.5cm} \includegraphics[width=4.5cm]{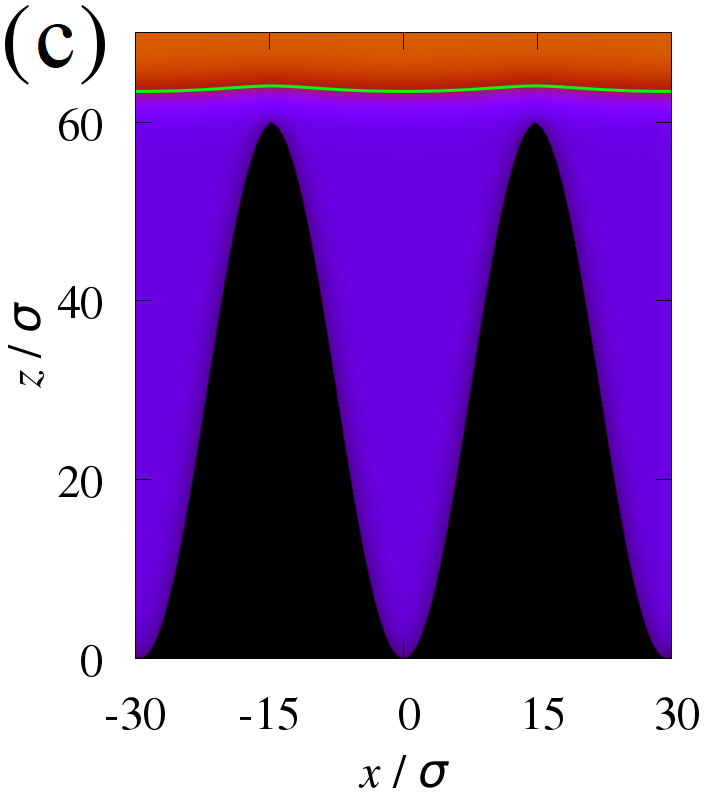} \hspace{0.2cm} \includegraphics[width=1.5cm]{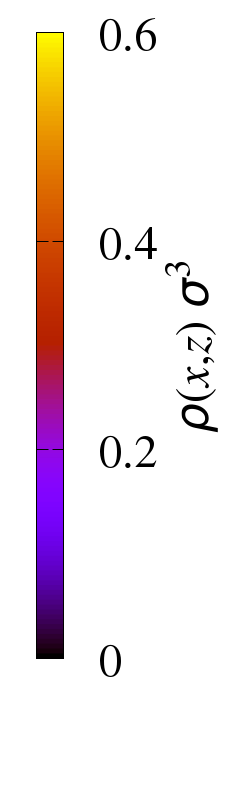}
\caption{Numerical DFT results for the equilibrium density profiles for the sinusoidal hard wall with $L=30\,\sigma$ and $A=30\,\sigma$  illustrating
a) microscopic, pre-osculation regime, b) macroscopic, geometry-dominated regime and c) microscopic, interface unbinding regime which correspond to
the three points highlighted in Fig.~\ref{regimes}. The green line denote the equilibrium height, $\ell(x)$, of the liquid-gas interface. }
\label{dens_profs}
\end{figure*}

\begin{figure}
\includegraphics[width=8cm]{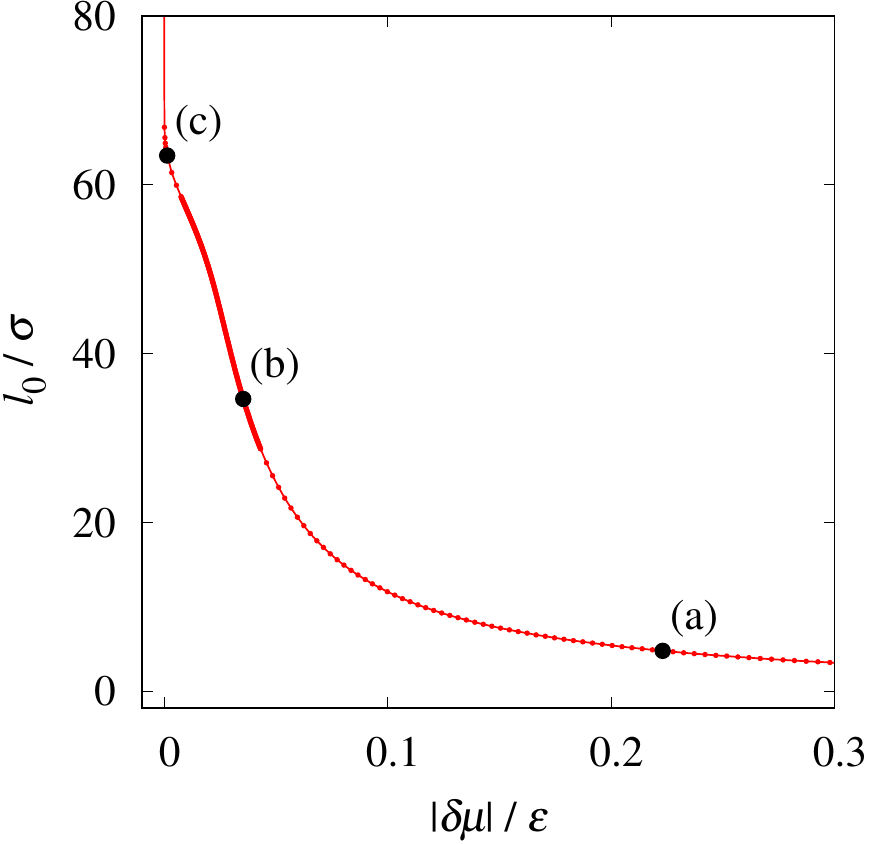}
\caption{Growth of the height of the liquid-gas interface, $\ell_0$, measured above the wall bottom, for the sinusoidal hard wall with $L=30\,\sigma$
and $A=30\,\sigma$. The points a), b) and c) correspond to the density profiles shown in Fig.~\ref{dens_profs}  representative of the three
adsorption regimes.} \label{regimes}
\end{figure}

\subsection{DFT results}

\subsubsection{Upton sum rule}

The contact sum rule for a planar hard wall in contact with a bulk fluid (i.e. in a semi-infinite geometry) elegantly relates the fluid density at
contact, $\rho_w$, to the bulk pressure \cite{hend}:
 \bb
 \beta p=\rho_w\,. \label{sumrule}
 \ee
For non-planar hard walls, the contact density becomes position-dependent and the theorem generalizes to \cite{upton}:
 \bb
 \beta p=\bar\rho_w\,. \label{upton_t}
 \ee
Here, $\bar\rho_w$ is the geometrically averaged contact density defined as
 \bb
  \bar\rho_w=\frac{1}{A_w}\int\rho({\bf s})ds\,,
 \ee
where  the integration is over the wall surface of total area $A_w=\int ds$. Here, $\rho({\bf s})$ is the value of the contact density at position
$x$ with ${\bf s}=(x,\psi(x))$ and $ds=dx\sqrt{1+\psi'(x)^2}$ is the local area element.

Being exact, the contact theorem is a useful test of the  accuracy of the numerical methods used to solve Eq.~(\ref{selfconsistent}). Errors arise in
a number of ways including the discretization used in the Piccard iteration and also from finite-size effects, since we have to impose that the
density takes its bulk value at some finite distance (which we set to be $50\,\sigma$) above the crests. In Fig.~\ref{upton-fig} we plot the
numerically determined value of $\beta p/\bar{\rho_w}$ as a function of the chemical potential for two different wall amplitudes and periods which we
use to study the meniscus osculation transition and adsorption isotherms. For the shallower geometry, with $L=100\,\sigma$ and $A=20\,\sigma$ there
is a very good agreement with the prediction of the Upton sum rule with the worst relative error about $2\%$. For the more corrugated wall with
$L=30\,\sigma$ and $A=30\,\sigma$ the agreement is still good but a little worse with the maximum relative error about $7\%$.

\subsubsection{Osculation transition and meniscus shapes}

We now turn our attention to the adsorption occurring on sinusoidal walls for a range of periodicities and amplitudes. All the DFT calculations were
performed at a temperature $T=0.925\,T_c$, where $T_c$ is the critical temperature of the bulk fluid ($k_BT_c=1.41\,\varepsilon$). At this
temperature, the corresponding densities of the coexisting bulk gas and liquid phases are $\rho_g=0.104\,\sigma^{-3}$ and
$\rho_l=0.431\,\sigma^{-3}$, respectively, and the chemical potential of saturation is $\mu_{\rm sat}=-3.943\,\varepsilon$.

At least qualitatively, the equilibrium density profile $\rho(x,z)$, and adsorption, fall into the three regimes as described earlier. This is
illustrated in Fig.~\ref{dens_profs} for the more corrugated wall where $L=30\,\sigma$ and $A=30\,\sigma$. In the density profile we also highlight
the local height of the liquid-gas interface, $\ell(x)$, which is shown as the green line. Of particular interest is the local height above the wall
bottom, $\ell_0$, whose dependence on the chemical potential is shown in Fig.~\ref{regimes}. Figures \ref{dens_profs} a)-c) and the corresponding
points on the adsorption of Fig.~\ref{regimes} correspond to the i) microscopic, pre-osculation regime, ii) macroscopic, geometry-dominated regime
and iii) microscopic, interfacial unbinding regime, respectively.

\begin{figure}
\includegraphics[width=9cm]{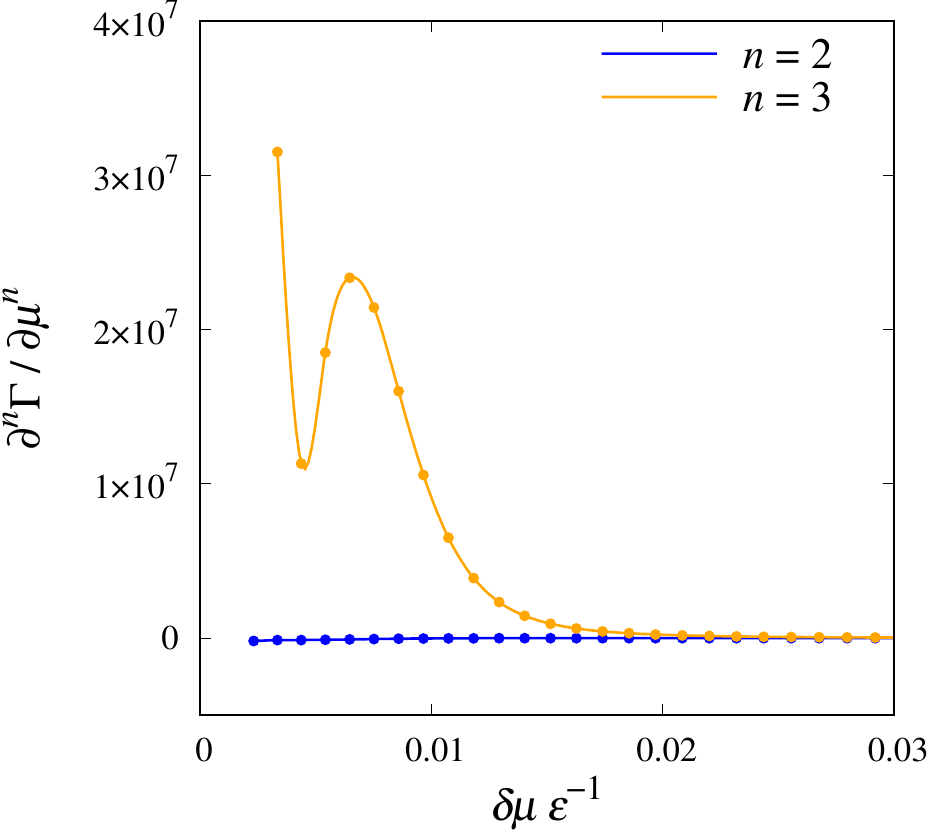}
\caption{Numerical DFT results comparing the second (blue line) and third (orange line) derivatives  of the excess adsorption $\Gamma$ w.r.t. $\mu$
for a sinusoidal hard wall with $L=100\,\sigma$ and $A=20\,\sigma$. The dramatic increase in the third derivative near
$\delta\mu/\varepsilon\approx0.01$ closely coincides with the growth of a meniscus near the bottom of the wall corresponding to a rounded meniscus
transition.} \label{numdiff-sin}
\end{figure}

Now we seek to be more quantitative.  Numerical evidence for a (rounded) osculation transition is shown in Fig.~\ref{numdiff-sin} for the shallower
geometry with $L=100\,\sigma$ and $A=20\,\sigma$ where we plot the second and third derivatives of the excess adsorption as a function of $\mu$. The
corresponding growth of the mid-point height $\ell_0$ is shown in Fig.~11 together with state points whose profiles we shall compare with the
predictions of the geometric construction. It can be seen from Fig.~10 that in contrast to the second derivative, $\partial^2\Gamma/\partial\mu^2$,
the third derivative, $\partial^3\Gamma/\partial\mu^3$,  undergoes a dramatic increase near $\delta\mu/\varepsilon\approx0.01$. This qualitative
difference between the second and third derivatives is consistent with the macroscopic, $5/2$ singularity predicted for the osculation transition --
recall Eq.~(\ref{gamma_osc_macro}). The sudden increase in the third derivative occurs close to the value of the chemical potential where the
meniscus lifts off the trough, see plots (3) and (4) in Fig.~12.

\begin{figure}
\centering
\includegraphics[width=8cm]{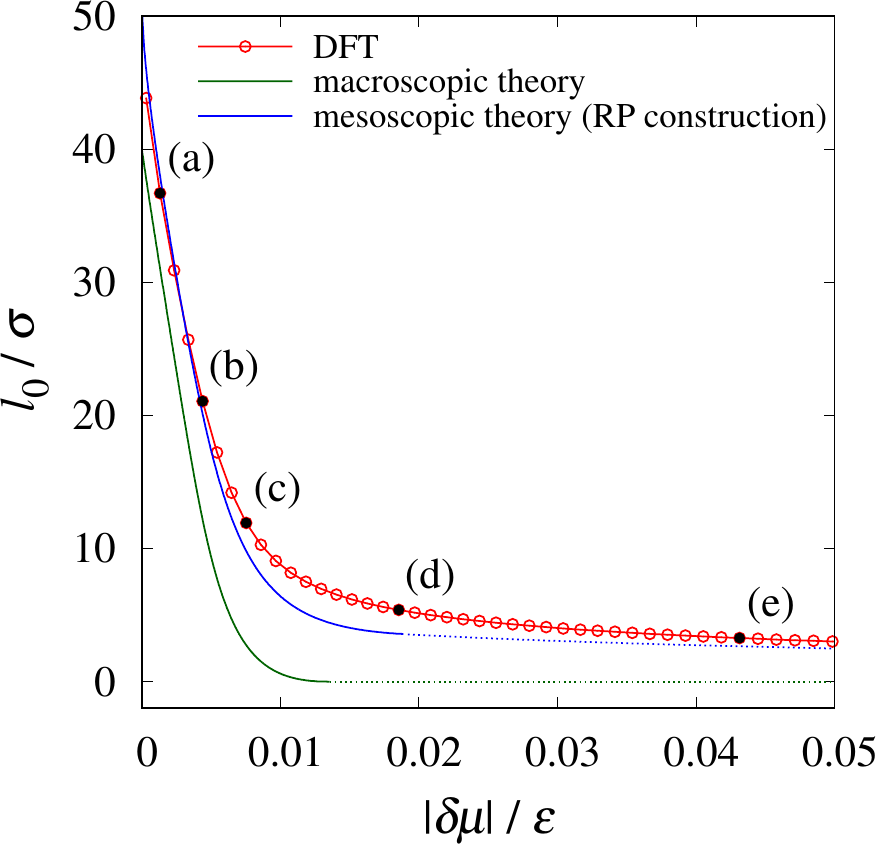}
\caption{Comparison of the numerical DFT results (red symbols) for $\ell_0$ as a function of $\delta\mu$ with the predictions of the purely
macroscopic theory (green line) and RP mesoscopic construction (blue line) for a sinusoidal hard wall with $L=100\,\sigma$ and $A=20\,\sigma$. The
continuation of the green and blue lines beyond the osculation transition are denoted by the respective dotted lines.} \label{numresults1}
\end{figure}

\begin{figure}
\includegraphics[width=8cm]{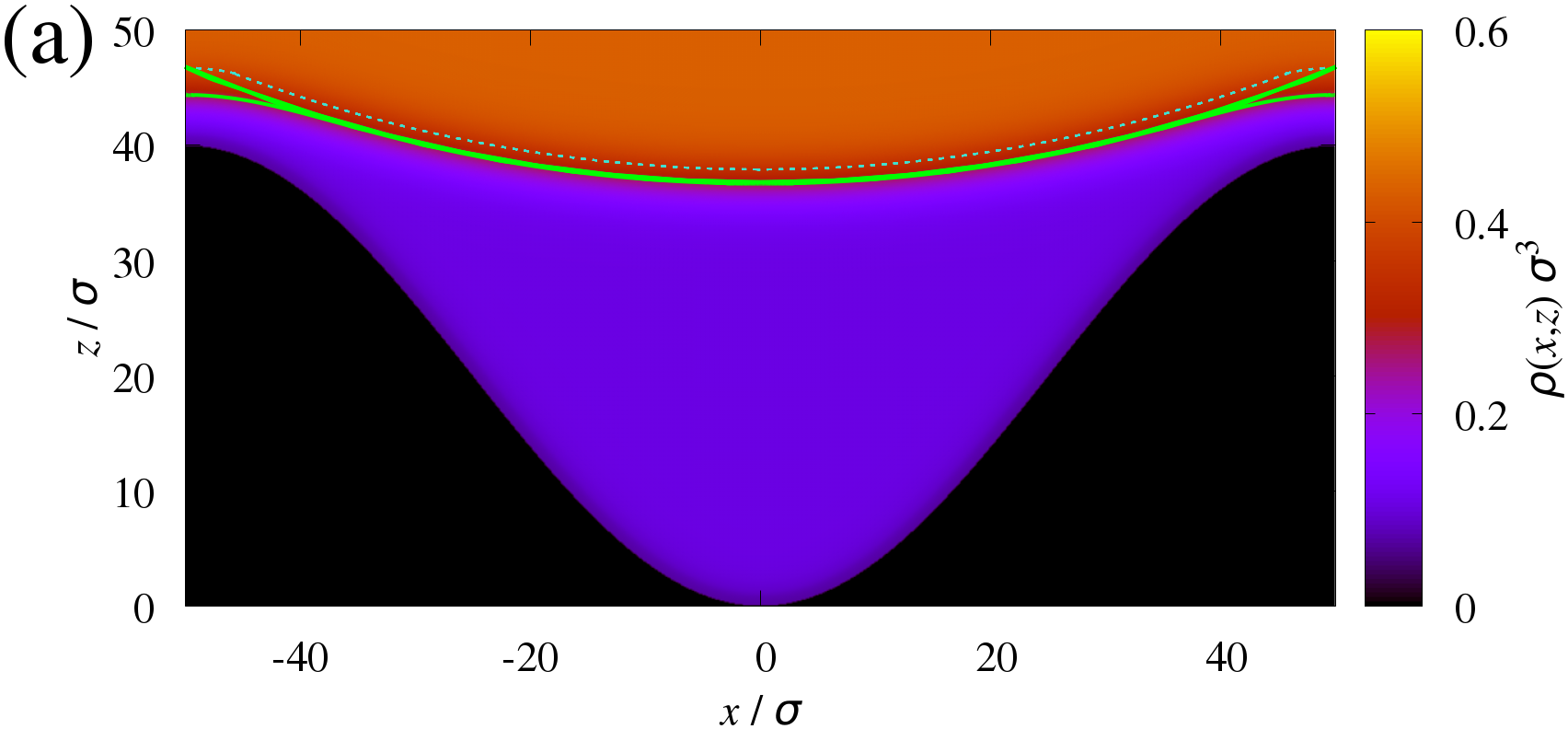} \\
\includegraphics[width=8cm]{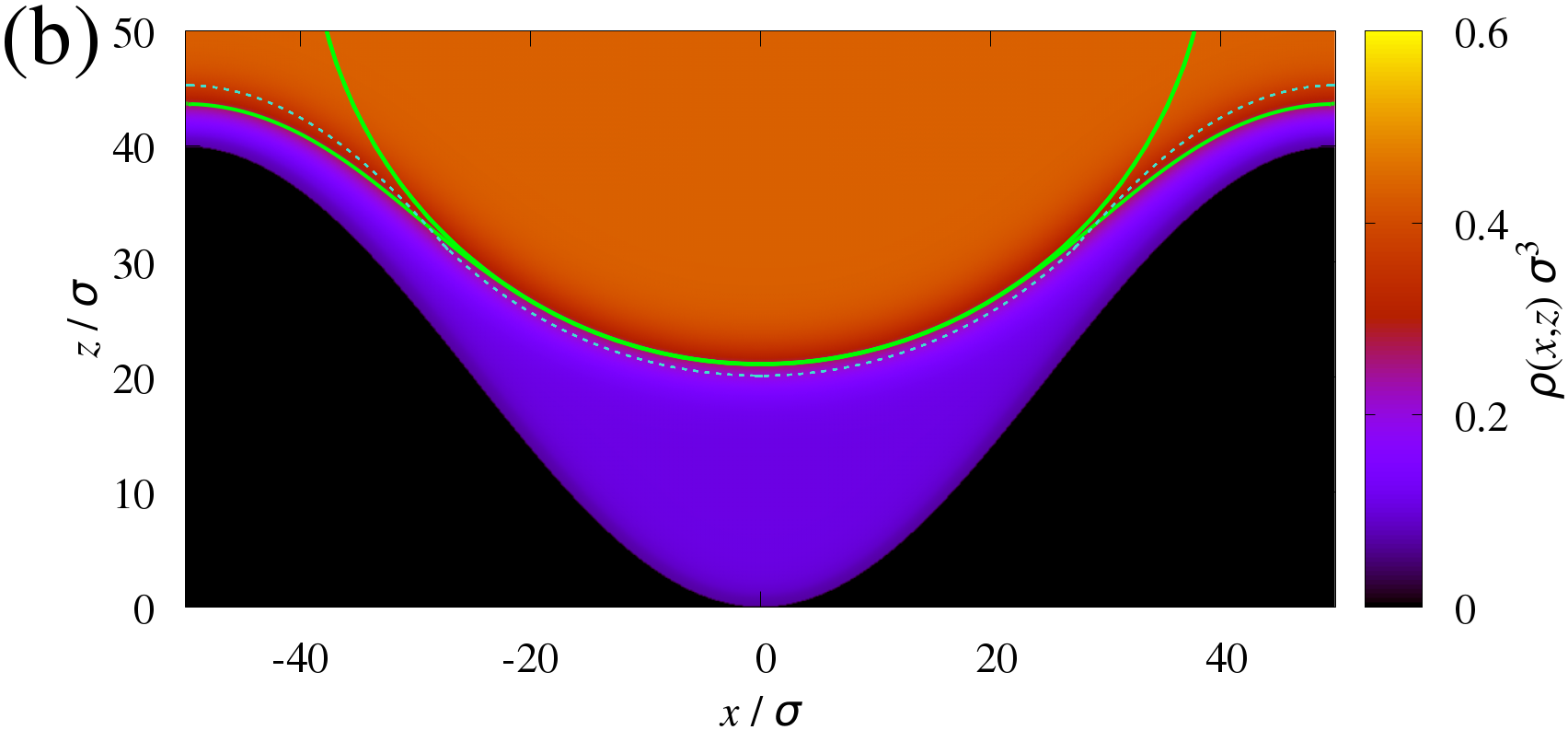} \\
\includegraphics[width=8cm]{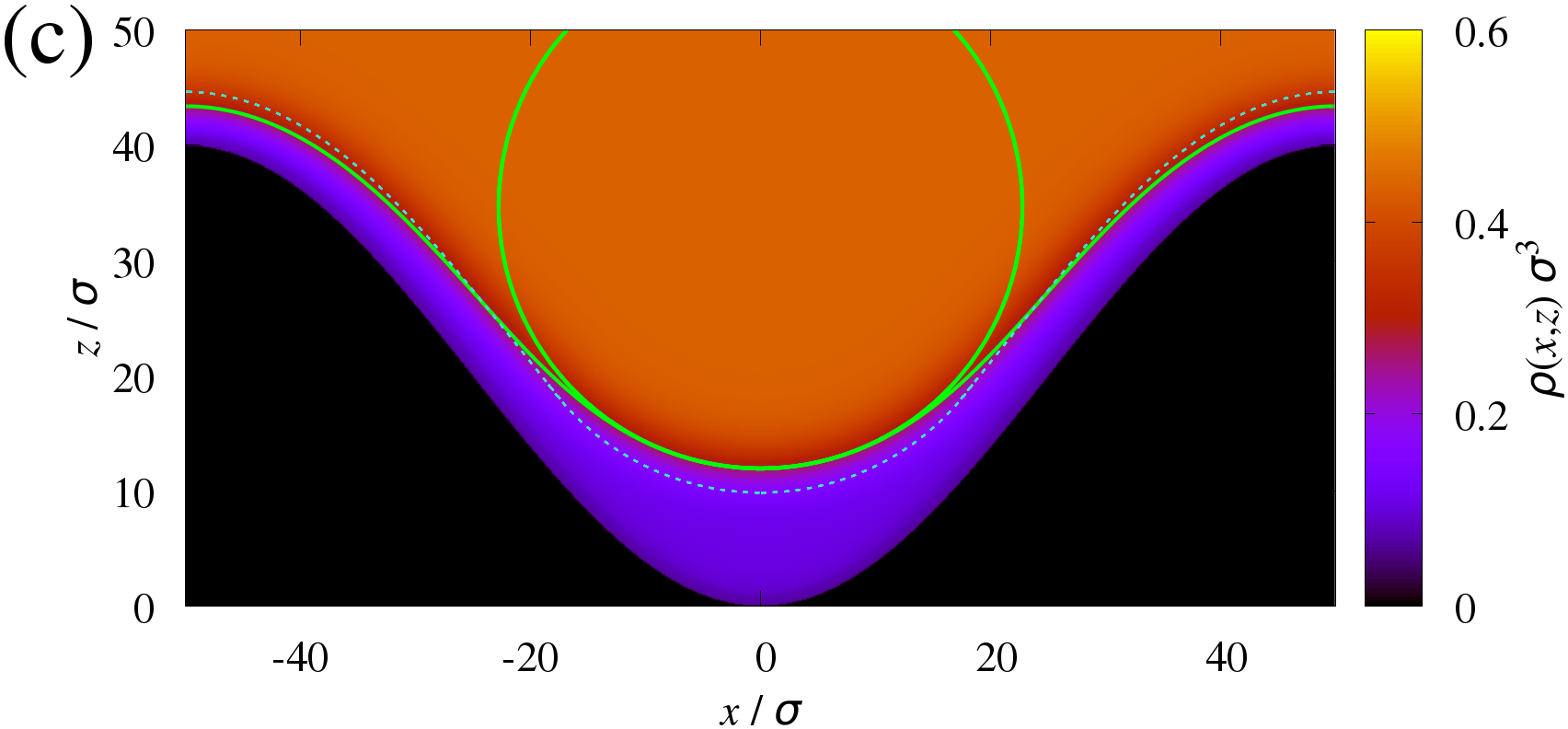} \\
\includegraphics[width=8cm]{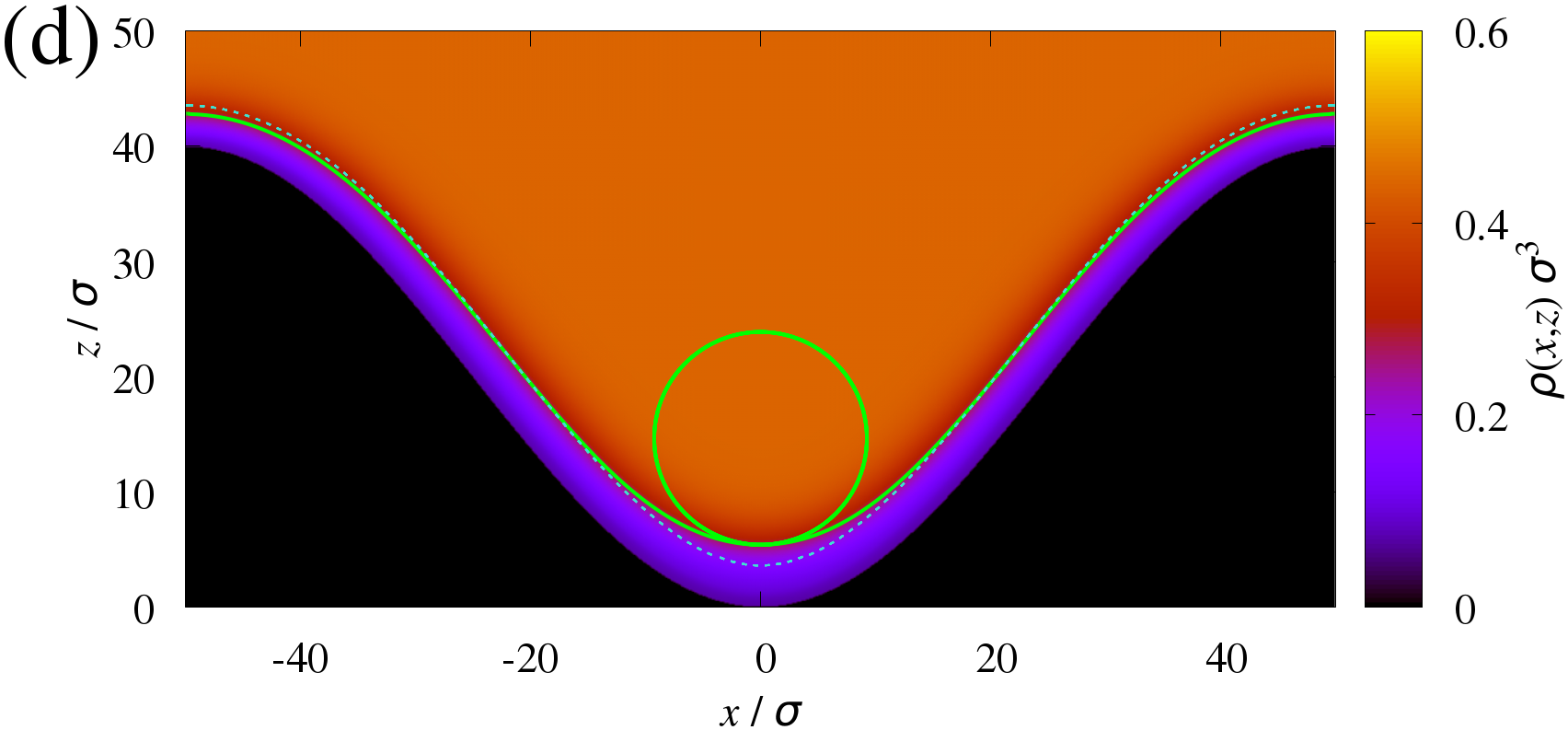} \\
\includegraphics[width=8cm]{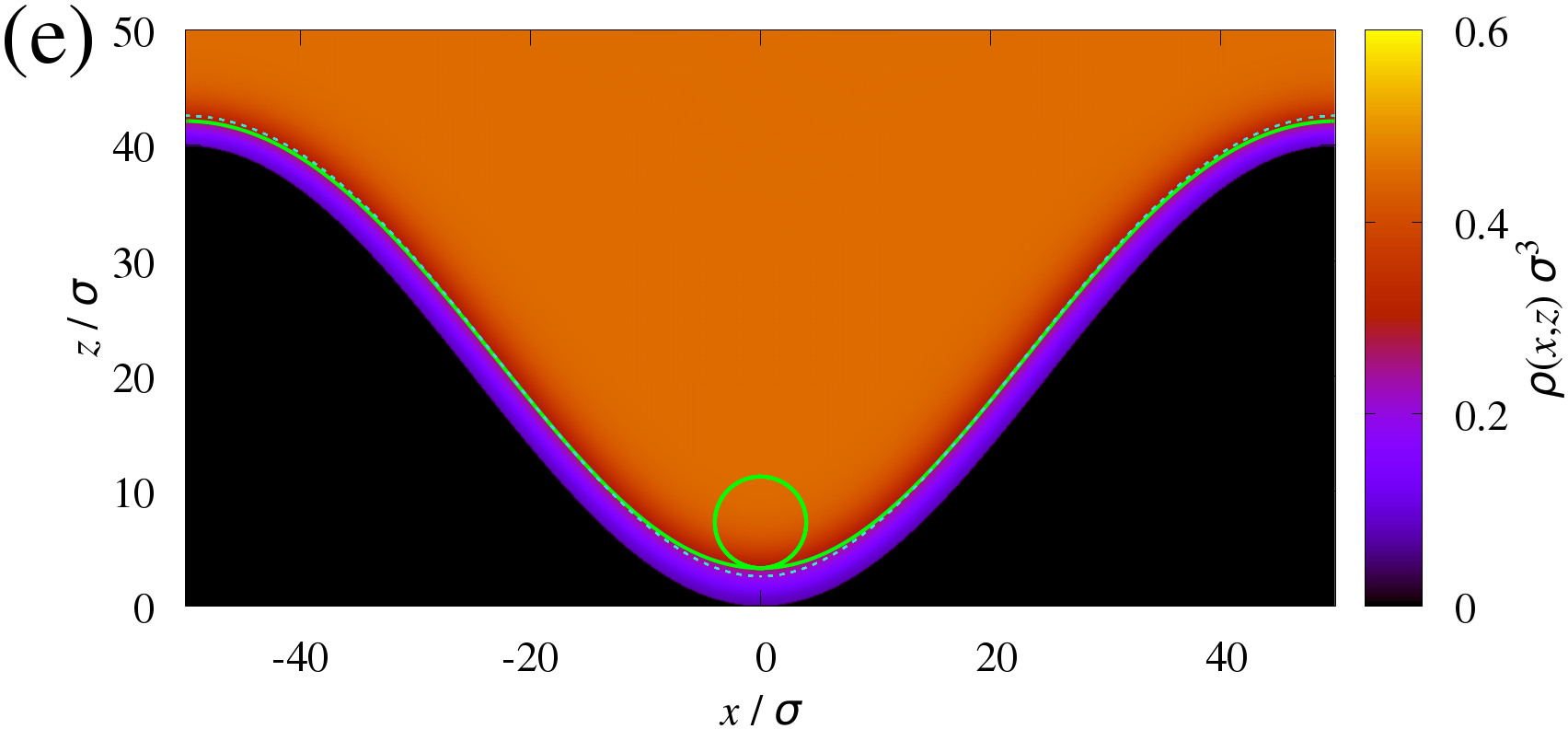}
\caption{Equilibrium DFT density profiles over a single period for the sinusoidal hard wall with the periodicity $L=100\,\sigma$ and amplitude
$A=20\,\sigma$ corresponding to the points in  Fig.~\ref{numresults1}. For each of these profiles three curves are highlighted. The solid green curve
is the DFT result for the location of the interface, $\ell(x)$, defined as the locus where the local fluid density is $(\rho_g+\rho_l)/2$. This is
compared with two other curves. The solid green circle has the Laplace radius $R$ and has been placed so that its lowest point coincides with the
true value of $\ell_0$. The green dashed line is the RP prediction for $\ell(x)$. } \label{profiles}
\end{figure}

\begin{figure*}
\includegraphics[width=5.5cm]{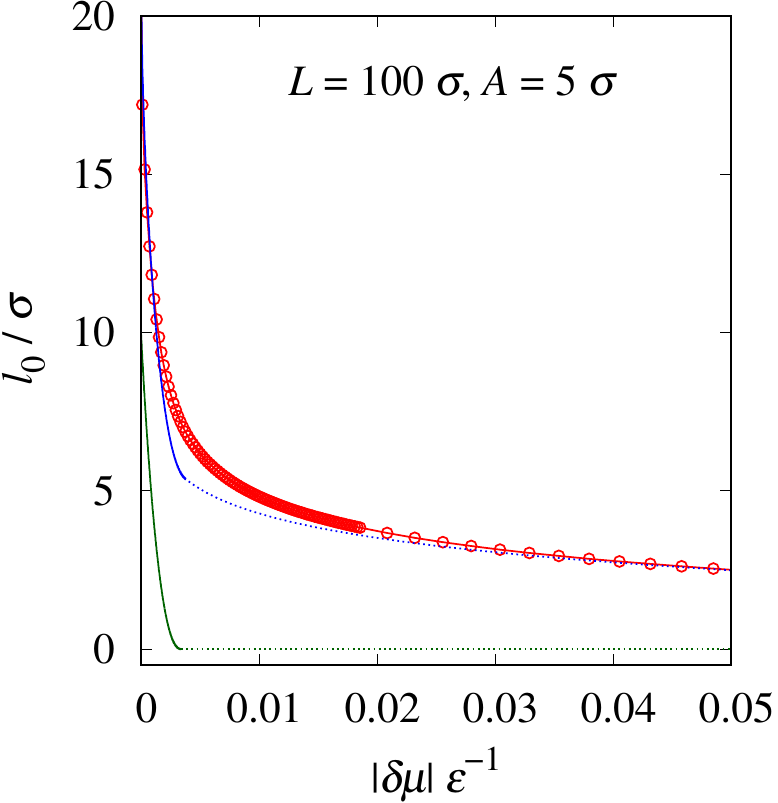}
\includegraphics[width=5.5cm]{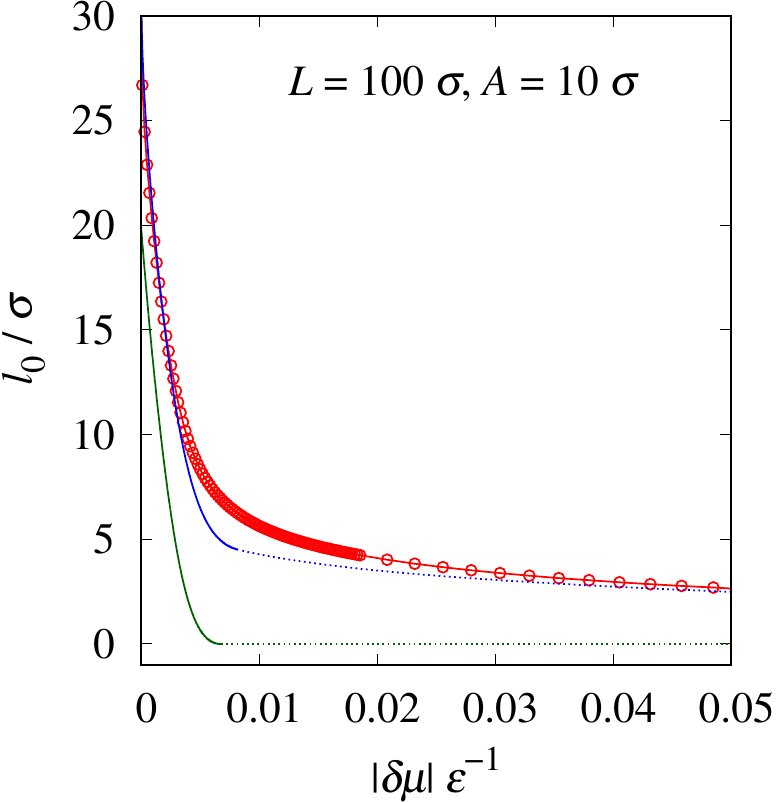}
\includegraphics[width=5.7cm]{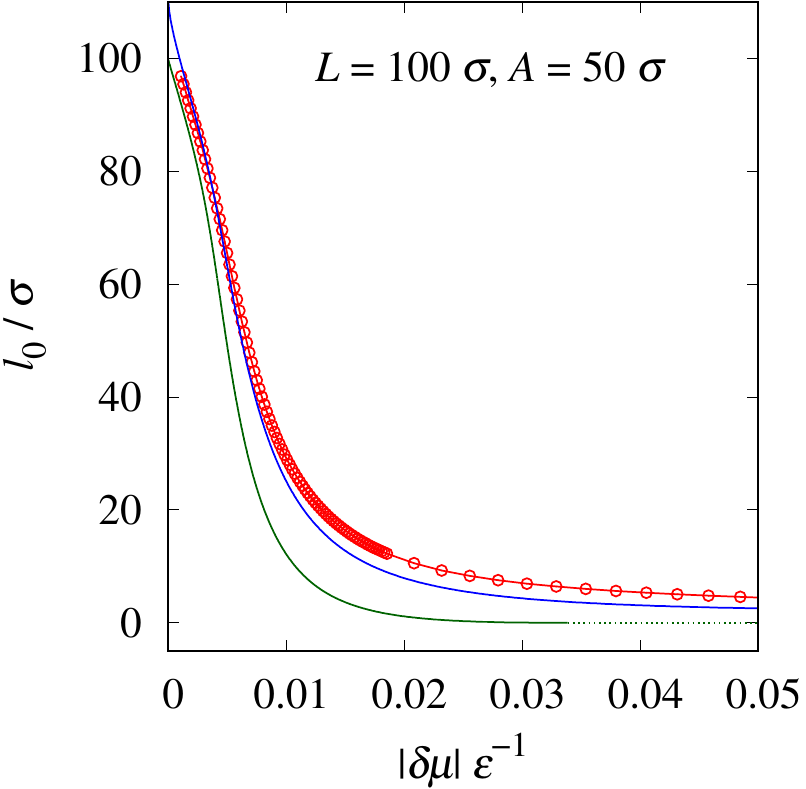} \\
\includegraphics[width=5.5cm]{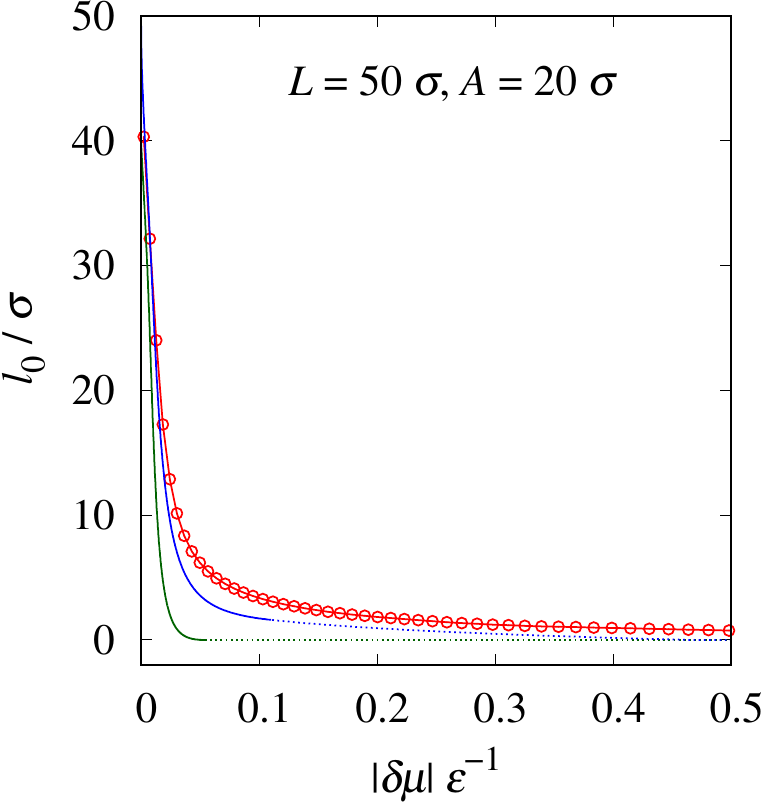}
\includegraphics[width=5.5cm]{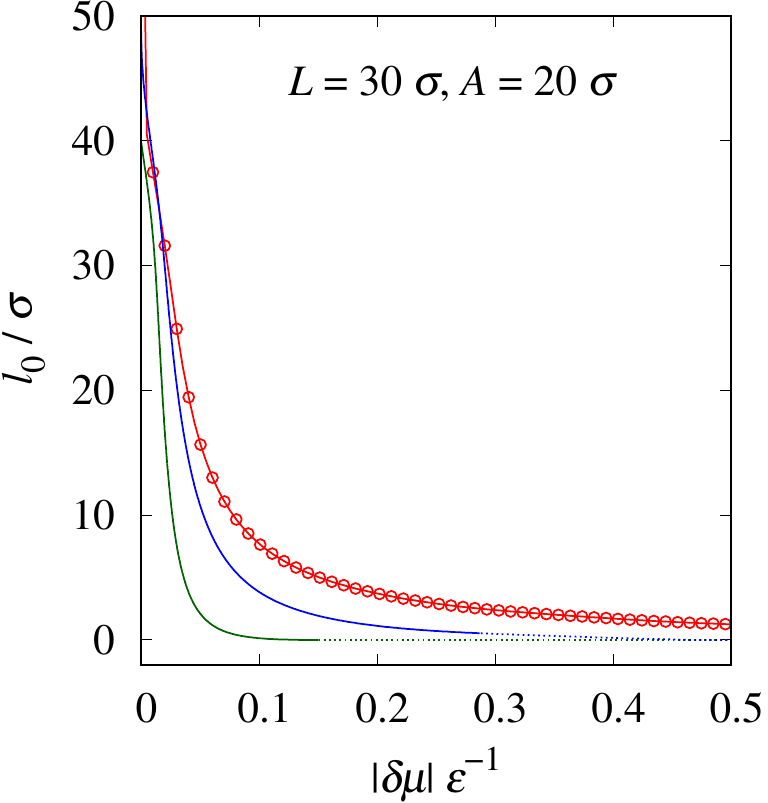}
\includegraphics[width=5.5cm]{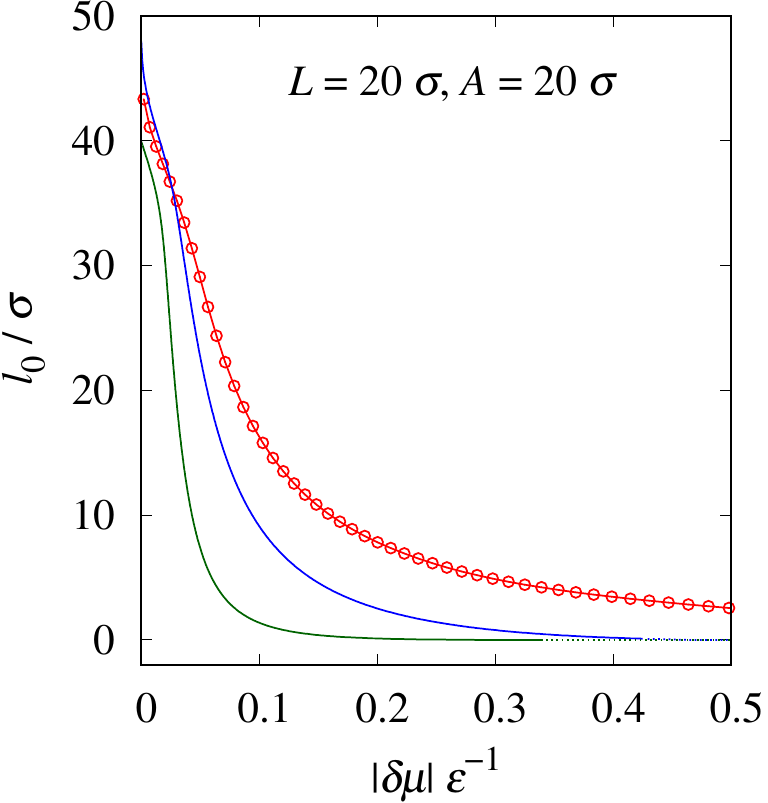}
\caption{Top row: Comparison of the meniscus height, $\ell_0$, as a function of $\delta\mu$ obtained from DFT (red symbols),  the macroscopic theory
(green line) and the mesoscopic RP construction (green line) for a sinusoidal hard walls with the fixed  periodicity $L=100\,\sigma$ and varying
amplitude: $A=5\,\sigma$ (left panel), $A=10\,\sigma$ (middle panel), and $A=50\,\sigma$ (right panel). Bottom row: The same comparison but for hard
sinusoidal wall of fixed amplitude $A=20\,\sigma$ and varying periodicity: $L=50\,\sigma$ (left panel), $L=30\,\sigma$ (middle panel), and
$L=20\,\sigma$ (right panel). Note that the scale of the abscissa in the bottom row is an order of magnitude larger than in the top row. }
\label{numresults2}
\end{figure*}

Next, we focus on the geometry dominated regime and test the macroscopic and mesoscopic predictions of section \ref{regime2} for the shape and height
of the meniscus. In Fig.~\ref{numresults1},  we compare the growth of the meniscus height, $\ell_0$, with the macroscopic prediction,
Eq.~(\ref{ell0}), and its mesoscopic correction following the RP construction. The purely macroscopic prediction is only in semi-quantitative
agreement with the DFT results.  The predicted height $\ell_0$ is systematically lower than the DFT results, which suggests that the discrepancy is
mainly due to the presence of thin drying layers which are allowed for, approximately, in the RP construction. The accuracy of this construction for
the meniscus shape is illustrated in Fig.~\ref{profiles} for five representative DFT density profiles corresponding to the thermodynamic points
highlighted in Fig.~\ref{numresults1}. In these figures we can compare the true shape of the interfacial profile $\ell(x)$ (solid green line) with
the prediction of the RP construction (green dashed line). The RP construction gives significantly better agreement with the DFT results, see the
blue line in Fig.~\ref{numresults1}. The RP construction is particularly accurate for the state points (1) and (2) where the meniscus is higher up
the trough, meeting the wall above the points of inflection of the sinusoid. It is probably not coincidental that the comparison between the DFT
results and the prediction of the RP construction is worst near the osculation transition. In these plots we have also highlighted a green circle of
Laplace radius $R$ which we have purposely placed to touch the \emph{true} value of $\ell_0$. It is apparent that the interfacial shape $\ell(x)$ is
always well described by the arc of the circular meniscus in the central region of the trough. However, it is clear from these plots that the simple
RP scheme of first coating the wall with the uniform layer of $\ell_\pi$ slightly overestimates the adsorption at the crests and slightly
underestimates the true interfacial height above the troughs. It might be possible to modify the RP construction by allowing for some local geometric
dependence to the wetting that we first coat the wall with before inserting the arc of the circular Laplace meniscus.

Further comparisons between the purely macroscopic and the RP predictions with the DFT results are presented in Fig.~\ref{numresults2} for various
wall parameters. In the upper row, we vary the wall amplitude for the fixed periodicity ($L=100\,\sigma$), while in the bottom row the periodicity is
varied for the fixed amplitude ($A=20\,\sigma$). In general, the agreement between the mesoscopic theory and the DFT is close but somewhat
deteriorates as the ratio $A/L$ increases. This is most likely because for highly curved walls any small discrepancy in $\ell_\pi$ or $R$ has a
significant impact on the resulting value of $\ell_0$. It appears that for sinusoidal walls the condition $A=L$ represents the limit beyond which
even the mesoscopic RP construction ceases to be quantitatively reliable.

\section{Summary}

In this paper we have studied the adsorption of fluids at smoothly corrugated completely wet walls using macroscopic theory, mesoscopic scaling
theory and microscopic DFT. In particular, we have focused on a rounded meniscus osculation transition occurring near the trough of the sinusoid
which is associated with the appearance of a meniscus as the chemical potential is increased towards bulk saturation. Macroscopically, this
transition is of $7/2$ order but is rounded due to the influence of a thin wetting layer arising from microscopic interactions. The scaling theory
that we developed for this predicts a non-trivial relationship between the interfacial height at the osculation and the wetting layer thickness. The
simple RP construction for mesoscopic corrections within the geometry-dominated regime is tested in our DFT where we show that it accurately
describes the adsorption on the corrugated surface, particularly where the meniscus is above the points of infection of the wall. For a sinusoidal
wall the adsorption falls into three regimes -- the pre-osculation regime, where the adsorption is determined by the microscopic interactions, a
post-osculation regime where the meniscus sits within the troughs of the wall and finally a complete wetting regime where the interface unbinds from
the crests. This final regime can again be only understood by taking into account the microscopic interactions. Our work can be extended in many
ways. In particular, it will be necessary to test the prediction for the value of the osculation exponent $\beta_{\rm osc}=3/7$ in systems with
dispersion forces and also in 2D; these studies are in progress. Also, recall that for meniscus depinning the order of the phase transition changes
when the walls are partially wet. This may also occur for meniscus osculation where the phase transition also competes with wetting and unbending
transitions \cite{rascon}. The surface phase diagram may be very rich in these systems similar to prediction for wetting on chemically patterned
surfaces. Returning to the case of complete wetting, our DFT results are suggesting that one may modify and improve the RP construction by allowing
curvature corrections to the thin wetting layer that first coats the wall. Finally, it would be interesting to examine in more detail how the shape
of the wall influences the divergences of the adsorption due to the unbinding of the liquid-gas interface as bulk coexistence is approached. To
correctly understand this, it is necessary to model the effect of interaction between the interface and the wall using the fully non-local binding
potential \cite{nonlocal}.

\begin{acknowledgments}
\noindent This work was financially supported by the Czech Science Foundation, Project No. 20-14547S.
\end{acknowledgments}

\end{document}